\documentclass[12pt, letterpaper]{article}\usepackage[]{graphicx}\usepackage[]{color}
\makeatletter
\def\maxwidth{ %
  \ifdim\Gin@nat@width>\linewidth
    \linewidth
  \else
    \Gin@nat@width
  \fi
}
\makeatother

\definecolor{fgcolor}{rgb}{0.345, 0.345, 0.345}

\usepackage{framed}
\makeatletter
 {\par\unskip\endMakeFramed%
 \at@end@of@kframe}
\makeatother

\definecolor{shadecolor}{rgb}{.97, .97, .97}
\definecolor{messagecolor}{rgb}{0, 0, 0}
\definecolor{warningcolor}{rgb}{1, 0, 1}
\definecolor{errorcolor}{rgb}{1, 0, 0}
\newenvironment{knitrout}{}{} 

\usepackage{alltt}
\usepackage[]{graphicx}
\usepackage[]{color}
\usepackage[comma]{natbib}
\usepackage{hyperref} 
\usepackage[]{mathalfa} \usepackage[]{amsfonts} \usepackage[]{amsmath} \usepackage{dsfont} \usepackage{amssymb} 
\usepackage{authblk} 
\usepackage[final]{changes} 
\usepackage{relsize} 
\usepackage{url} 
\usepackage{enumitem} 
\usepackage[noframe]{showframe} 
\newcommand{\E}{\operatorname{E}}
\newcommand{\Var}{\operatorname{Var}}
\newcommand{\DELTA}[1][i]{\beta x_{#1}+\kappa x_{#1}^2+\epsilon_{#1}}
\allowdisplaybreaks 

\newcommand{\blind}{1}

\IfFileExists{upquote.sty}{\usepackage{upquote}}{}
\begin{document}

\def\spacingset#1{\renewcommand{\baselinestretch}%
{#1}\small\normalsize} \spacingset{1}

\if1\blind
{
  \title{\bf A hierarchical model of non-homogeneous Poisson processes for Twitter retweets}
  \author[1,2]{Clement Lee}
  \author[1]{Darren J Wilkinson}
  \affil[1]{School of Mathematics, Statistics and Physics, Newcastle University, UK}
  \affil[2]{Open Lab, School of Computing, Newcastle University, UK}
  \maketitle
} \fi

\if0\blind
{
  \bigskip
  \bigskip
  \bigskip
  \begin{center}
    {\LARGE\bf A hierarchical model of non-homogeneous Poisson processes for Twitter retweets}
\end{center}
  \medskip
} \fi

\bigskip
\begin{abstract}
  We present a hierarchical model of non-homogeneous Poisson processes (NHPP) for information diffusion on online social media, in particular Twitter retweets. The retweets of each original tweet are modelled by a NHPP, for which the intensity function is a product of time-decaying components and another component that depends on the follower count of the original tweet author. The latter allows us to explain or predict the ultimate retweet count by a network centrality-related covariate. The inference algorithm enables the Bayes factor to be computed, in order to facilitate model selection. Finally, the model is applied to the retweet data sets of two hashtags.
\end{abstract}

\vfill

\newpage
\spacingset{1.45}

\section{Introduction} \label{sect.intro}
Statistical modelling of online social media such as Twitter has become increasingly popular, because of the richness and availability of the data in temporal and topological aspects. As an introduction to the model for Twitter retweets proposed in this article, we will give a brief review for each of the two aspects.

\subsection{Temporal dynamics} \label{sect.intro_temporal}
A common approach to modelling temporal dynamics is the use of one-dimensional non-homogeneous Poisson Processes (NHPP). Specifically, if a sequence of events is assumed to arise from a NHPP with intensity function $h(t)\geq0$, the random variable of the number of events within the interval $[t_1,t_2]$ will follow a Poisson distribution with mean $\displaystyle\int_{t_1}^{t_2}h(t)dt$, and is independent of the random number of events in any other disjoint interval. The special case where $h(t)$ is constant over time is called the homogeneous Poisson process (HPP). 

Examples of using the HPP for Twitter data include \cite{som10}, \cite{pasc10}, \cite{klms14, klms15}, and \cite{mcn13}, but it usually does not describe data realistically because it assumes the interarrival times of events are independent and identically distributed (iid) exponential random variables. \cite{sl15} observe bursty dynamics and temporal fluctuations in tweets with hashtag \#\verb'ledebat', over the two-week period leading to the 2012 French presidential election, and show the departure of the data from one simulated from a HPP without fitting a stochastic process. It is therefore natural that the more general NHPP, or extensions thereof, are more often used in the literature. For example, in \cite{smtwe11}, a NHPP is being used for semi-supervised detection of an anomaly in pollution-related tweets.

Quite often the intensity function $h(t)$ is specifically designed or chosen to capture temporal patterns observed in the data. For example, \cite{maaj13} and \cite{mm15} both fit a NHPP to the occurrences of international brand names on Twitter, which exhibit strongly correlated user behaviour and bursty collective dynamics over time. The former incorporate long range temporal correlations in $h(t)$, resulting in interarrival times that are marginally distributed according to the power law, while the latter consider $h(t)$ as a product of stochastic global user interest and approximately deterministic user activity over time. Such a way of splitting $h(t)$ into two components is also seen in \cite{swsb14}, and \cite{mmnb17}. The former fit a NHPP to the popularity dynamics of Twitter hashtags and Physical Review papers, where $h(t)$ is a product of a decreasing function of time and a term increasing linearly with the number of events, intended to capture the effect of how the attractiveness of an individual item ages, and the effect of preferential attachment, respectively. The latter fit a NHPP to the retweets of popular Twitter users with $h(t)$ proportional to the product of $t^{-\lambda}$ and $e^{-\theta t}$, and attempt to explain the two components by a decision-based queueing process, rather than preferential attachment, and loss of interest over time, respectively. The NHPP with this specific deterministic form of $h(t)$ is termed the \textit{hybrid process}, which will be the cornerstone of our proposed model.

While relaxing the stationarity assumption of the HPP leads to the NHPP, the former, as one of the simplest point processes, has many other properties (and therefore has alternative characterisations), such as the interarrival times being independent and identically distributed (i.i.d.) exponential random ariables, the independence of the random variables of the numbers of points in disjoint intervals, and that points are i.i.d. over a bounded region of time or space conditional on the total number of points in that region. This allows defining broader classes of point processes by generalising the HPP in various ways. For example, relaxing the exponential distribution assumption for the interarrival times gives rise to the renewal processes. Further allowing that the interarrival times form a Markov chain leads to the Wold process. Alternative joint distributions being specified for the points in a bounded domain given the total number of points leads to finite point processes. All these processes can be found in, for example, \cite{dv03}.

Apart from generalising the HPP in various ways, extensions can be made for point processes in general. For example, if there are measurements (marks) associated with the locations of the points, the observed data can be modelled by a marked point process, in which the locations are modelled by a point process and the marks are further modelled by a distribution, possibly conditional on the locations. Also, while the (first-order) intensity function characterises the NHPP, higher-order intensity functions, which are defined in a similar way, can be used to complement the characterisation of other processes, such as the Markov point process, also known as the Gibbs process. For further references of these processes, please see, for example, \cite{dv03} and \cite{diggle13}.

Another common approach to extending a NHPP is the incorporation of stochasticity in $h(t)$. This means that events are assumed to arise from a NHPP \textit{conditional on $h(t)$}, which in turns arises from a separate stochastic process. One prominent example is the log-Gaussian Cox process \citep{msp98}. Regarding the models for Twitter data relevant to our research, one example is the aforementioned model by \cite{mm15}. \cite{pk11} use a Markov-modulated Poisson process, in which $h(t)$ varies according to a Markov process, in their application of identification and spatio-temporal analysis of topics on Twitter. \cite{bsjc15} propose a self-excited Hawkes process (SEHP), in which $h(t)$ jumps simultaneously when an event occurs and decays before the next event occurs, and argue that such their model outperforms the one by \cite{swsb14}, in terms of prediction accuracy, when applied to the same set of data.

The SEHP is being used widely in different fields to capture triggering and clustering behaviour \citep{reinhart18}. One similar context to modelling Twitter data is earthquake modelling in seismology. \cite{ogata88} developed the temporal epidemic-type aftershock sequence (ETAS) model, in which mainshocks arise from a NHPP called the background process, and aftershocks arise from a different NHPP triggered by the occurrence of a mainshock or an aftershock. While the Twitter original tweets and retweets can be seen as analogous to the mainshocks and aftershocks in the ETAS model, respectively, subtle differences to our proposed model exist and will be explained in Section \ref{sect.model}. The spatio-temporal version of the ETAS model is reviewed in \cite{ca17}, with a focus on inference approaches.

One final aspect of extending a temporal NHPP or general point process is its spatial counterpart or spatio-temporal generalisation. The spatial component exists in some of the references above, such as \cite{reinhart18} for the SEHP, but is not reviewed extensively here due to the temporal nature of our research. For further references, please see \cite{cressie93}, \cite{ipss08}, \cite{dv08}, \cite{gdfg10}, \cite{cw11}, \cite{diggle13} and \cite{brt16}.

\subsection{Topological Aspects} \label{sect.intro_network}
Twitter data usually comes with information such as number of followers or even who follows whom, that is, the directed edges in the user network, therefore enabling modelling of its social network, static or dynamic. For example, \cite{bsz15} collect tweets of specific topics associated with competing hashtags, and observe the departure of the degree distribution of the retweet network from one predicted by the classical preferential attachment model \citep{ba99}. They propose a variant called the Superstar model, in which a vertex enters the network by connecting to either the lone superstar, with the same probability across all vertices, or the rest of the network otherwise, according to original preferential attachment rule.

Whenever the data permits, it is natural to extend a model to account for the temporal dynamics and the network structure simultaneously, see, for example, \cite{llwz14}. \cite{xzw11} and \cite{wgzx11} build a framework for data on Sina weibo, a Chinese counterpart of Twitter, that divides users into communities and models information generation, receiving, and processing and diffusion by the power law, a HPP, and a multiplicative model of individual reading habits and relation strength, respectively. It is also possible to model network structure and information diffusion simultaneously without using time as a dimension. Both \cite{llxw12} and \cite{ntomtkm16} use a Galton-Watson branching process model, for data of video contents shared on online social networks, and reply trees in Twitter, respectively.

Usually and implicitly assumed in the models aforementioned is that the network, if concerned, remains unchanged throughout the observation period, which may be unrealistic for Twitter data given the ease of following other users. Therefore efforts have been made to model the dynamics of information diffusion and network evolution simultaneously, as it is natural to conjecture that they co-evolve over time. \cite{ad15} propose a tweet-retweet-follow model, which is characterised by events of a follower of a retweeter becoming also a follower of the original tweet author, conditional on the original tweet being created and retweeted. \cite{fetal15} consider the follower and the retweet adjacency matrices, and model the co-evolution through a system of dynamic equations of these two matrices. \cite{slbc17} assume no network evolution but incorporate the influence between users according to the network in a multivariate Hawkes process model, in which each user-topic pair has its own intensity function. \cite{lcb16} introduce a Twitter-Network topic model, which comprises a hierarchical Poisson-Dirichlet process model for the text and hashtags, and a Gaussian process based random function model for the followers network, and is applied to a data set of tweets with certain keywords.

Instead of modelling temporal and network dynamics merely according to some stochastic processes, one can look into how network summary measures, such as number of followers, and other variables affect either or both of them, thus identifying useful covariates for predictions for retweet behaviour and network influence. \cite{setal14} employ a negative binomial regression model for the retweet count, to investigate how message content/style and public attention to tweets relate to the retweet activity in a disaster. \cite{zxplz11} apply a logistic regression model to the data of whether a tweet is being retweeted from the point of view of a \textit{follower}. \cite{hdd13} include a regression part in their co-factorization machines, which are for discovering topics users are interested in. They suggest that both network measures and content are important in determining retweets. However, the relationships among the covariates are not reported in all three analyses, thus presenting the risk of potential collinearity and overfitting.

Commonly observed and modelled in the aforementioned literature is the power law phenomenon. Examples in temporal aspects include interarrival times \citep{glmf09, xzw11, wgzx11}, and tweet or retweet rate \citep{maaj13, mmnb17}, while examples in topological aspects include network degree \citep{llwz14} and size and depth of reply trees \citep{ntomtkm16}. Regarding network influence, the power law has been observed in retweet count \citep{hdd13}, count of in-links for blogs \citep{glmf09}, view count of videos \citep{mka17}, and citation count \citep{swsb14}. Interestingly, there have been no studies on the relationships among these variables following the power law.

\subsection{Proposed model} \label{sect.intro_ours}
The research reported in this article stemmed from investigating all tweets (both original and retweets) with two specific hashtags. Compared to the analysis by \cite{swsb14}, we dig one level deeper as we model the retweets by a collection of NHPPs conditional on the existence of the original tweets, simply called the originals hereafter. The specific NHPP used in the modelling is based on the hybrid process as in \cite{mmnb17}, but without resorting to discretising the data when it comes to inference. While it is straightforward to fit a hybrid process to the originals, as we will illustrate in Section \ref{sect.data}, the novelty of this article is the hierarchical modelling of retweets. Specifically, all retweets of each original are modelled by a NHPP, with a latent component in $h(t)$ that depends on the follower count (of the author of the original) and determines the ultimate retweet count. All the retweet processes are in turn enveloped in one single hierarchical model so that information can be pooled to estimate the parameters. 

There are a few merits of including follower count, which is essentially the in-degree of a user, and retweet count in the way described above, both of which are observed to follow the power law empirically. First, it presents a network centrality-related covariate as the potential driving force of retweet behaviour or network influence, while simultaneously modelling the temporal dynamics. Second, such a way of incorporating network summary measures enables us to capture any effect attributed to the power law phenomenon, while avoiding the overhead of an explicit network structure, which is usually computationally expensive to construct. Furthermore, directly using the follower count, which can vary over the observation period even for the same user, already partially accounts for the effect of network evolution over time. Finally, this model is generative in the sense that, conditional on the follower count of the authors of the originals (which can be easily generated by the power law), we can simulate a realistic process of processes, each of which corresponds to how retweets of a particular original grow over time.

The rest of the article is divided as follows. Introduced and explored in Section \ref{sect.data} are the two data sets, the subsets of which are fitted by the hybrid process and its special case. The hierarchical model for retweets is introduced in Section \ref{sect.model}, with its likelihood derived. The inference algorithm and the model diagnostics procedure are outlined in Sections \ref{sect.inf} and \ref{sect.gof}, respectively. The model is applied to the two previously introduced data sets in Section \ref{sect.app}, as well as a simulated data set in Section \ref{sect.sim} to show that the model is realistic and that the inference algorithm performs well. Section \ref{sect.con} concludes the article.

\section{Data and exploratory analysis} \label{sect.data}

A set of tweets (both originals and retweets) with the hashtag \#thehandmaidstale was collected on 2017-06-14 for 21 hours after one episode of the relevant TV series was broadcasted. There are in total 2043 originals, 265 of which have been retweeted at least once during the observation period. The times of the originals are plotted on the left of Figure~\ref{fig.both_hist} using a histogram, with bins of 10 minutes. For these 265 originals there are 971 retweets, while the most retweeted original, called the top original hereafter, has been retweeted 204 times. For each original, the cumulative retweet count is plotted over time on the left of Figure \ref{fig.both_times}, meaning that there are 265 trajectories of different colours in total.

To examine potentially different tweeting behaviour of different hashtags, a set of tweets with the hashtag \#gots7 was collected on 2017-07-16 for around 4.4 hours \textit{before} the 7$^{\text{th}}$ season premiere of the TV series Game of Thrones was broadcasted. The numbers of originals, originals retweeted, total retweets and retweets of the top original are provided in Table \ref{tab.summaries} alongside the counterparts for \#thehandmaidstale data reported above. The histogram of originals is plotted on the right of Figure \ref{fig.both_hist}, while the cumulative retweet counts over time for each original are plotted on the right of Figure \ref{fig.both_times}.

\begin{table}
  \centering
  \renewcommand{\arraystretch}{0.8}
  \begin{tabular}{|l|c|c|}
    \hline
    & \#thehandmaidstale & \#gots7 \\ \hline
    Originals & 2043 & 25420 \\ \hline
    Originals retweeted & 265 & 3145 \\ \hline
    Total retweets & 971 & 29751 \\ \hline
    Retweets of top original & 204 & 3204 \\ \hline
  \end{tabular}
  \caption{Summaries of originals and retweets for the two data sets.}
  \label{tab.summaries}
\end{table}

\begin{knitrout}
\definecolor{shadecolor}{rgb}{0.969, 0.969, 0.969}\color{fgcolor}\begin{figure}[htbp!]

{\centering \includegraphics[width=0.49\linewidth]{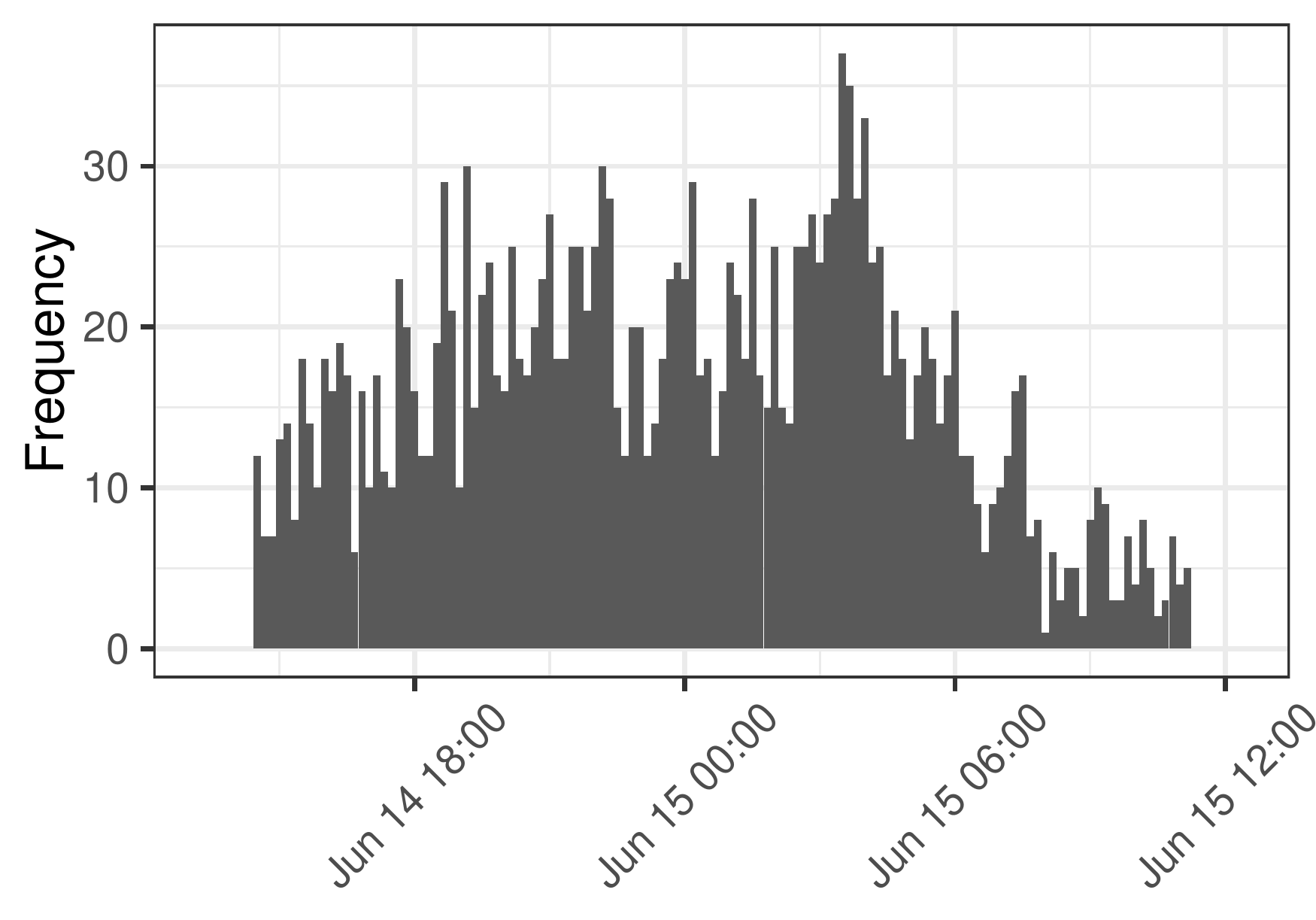} 
\includegraphics[width=0.49\linewidth]{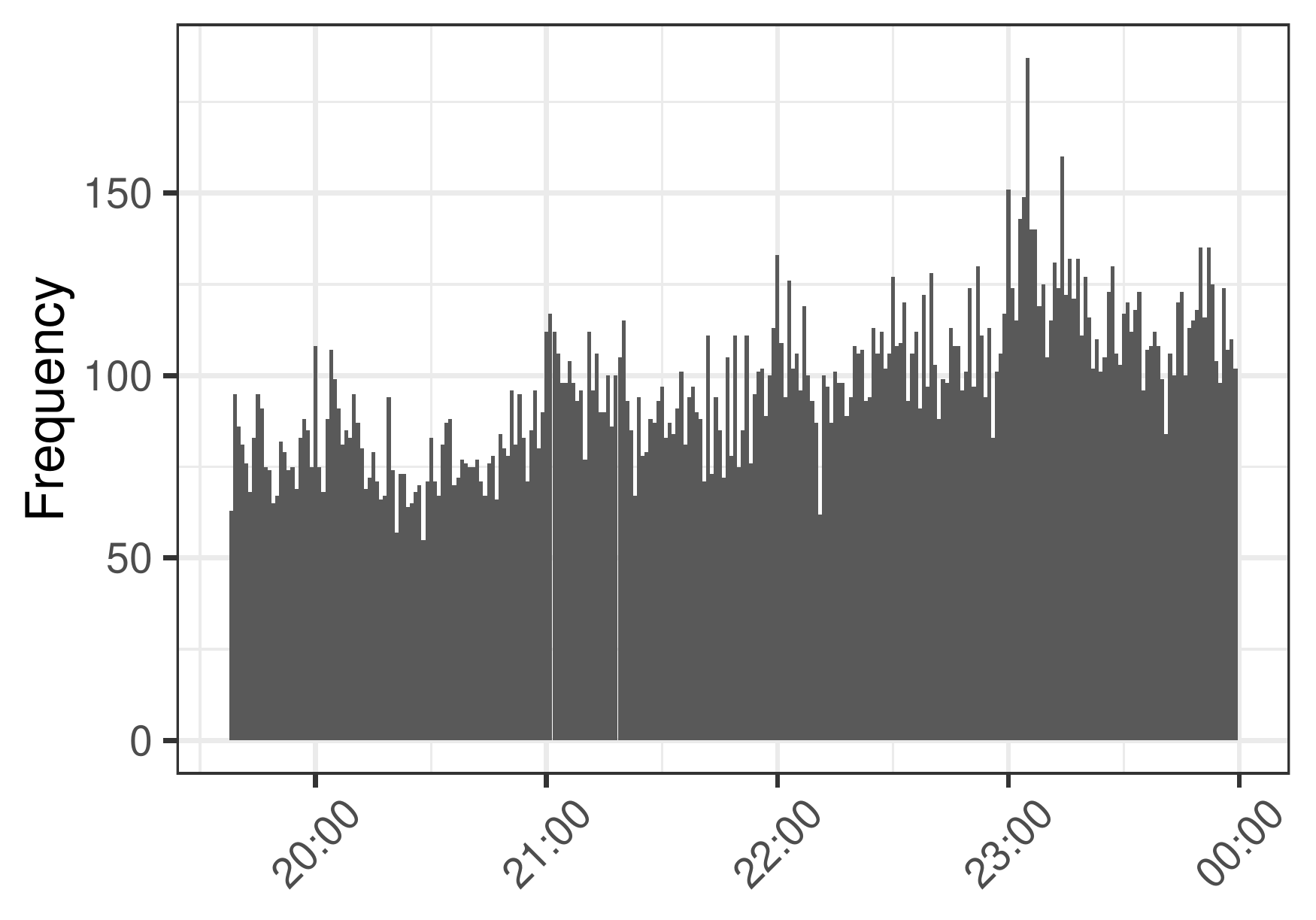} 

}

\caption[Histogram of originals over time for the \#thehandmaidstale (left) and \#gots7 (right) data]{Histogram of originals over time for the \#thehandmaidstale (left) and \#gots7 (right) data.}\label{fig.both_hist}
\end{figure}

\end{knitrout}

\begin{knitrout}
\definecolor{shadecolor}{rgb}{0.969, 0.969, 0.969}\color{fgcolor}\begin{figure}[htbp!]

{\centering \includegraphics[width=0.49\linewidth]{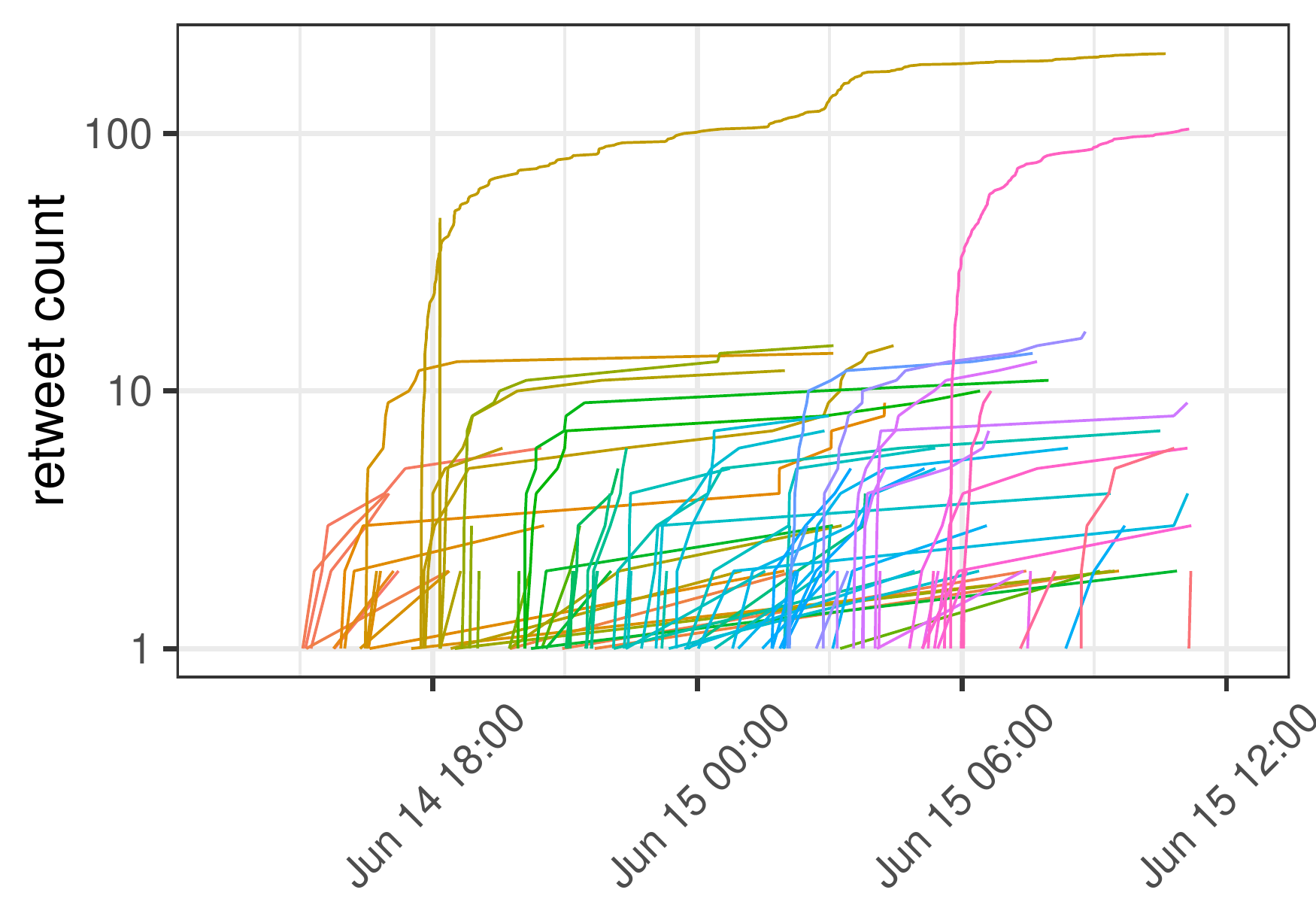} 
\includegraphics[width=0.49\linewidth]{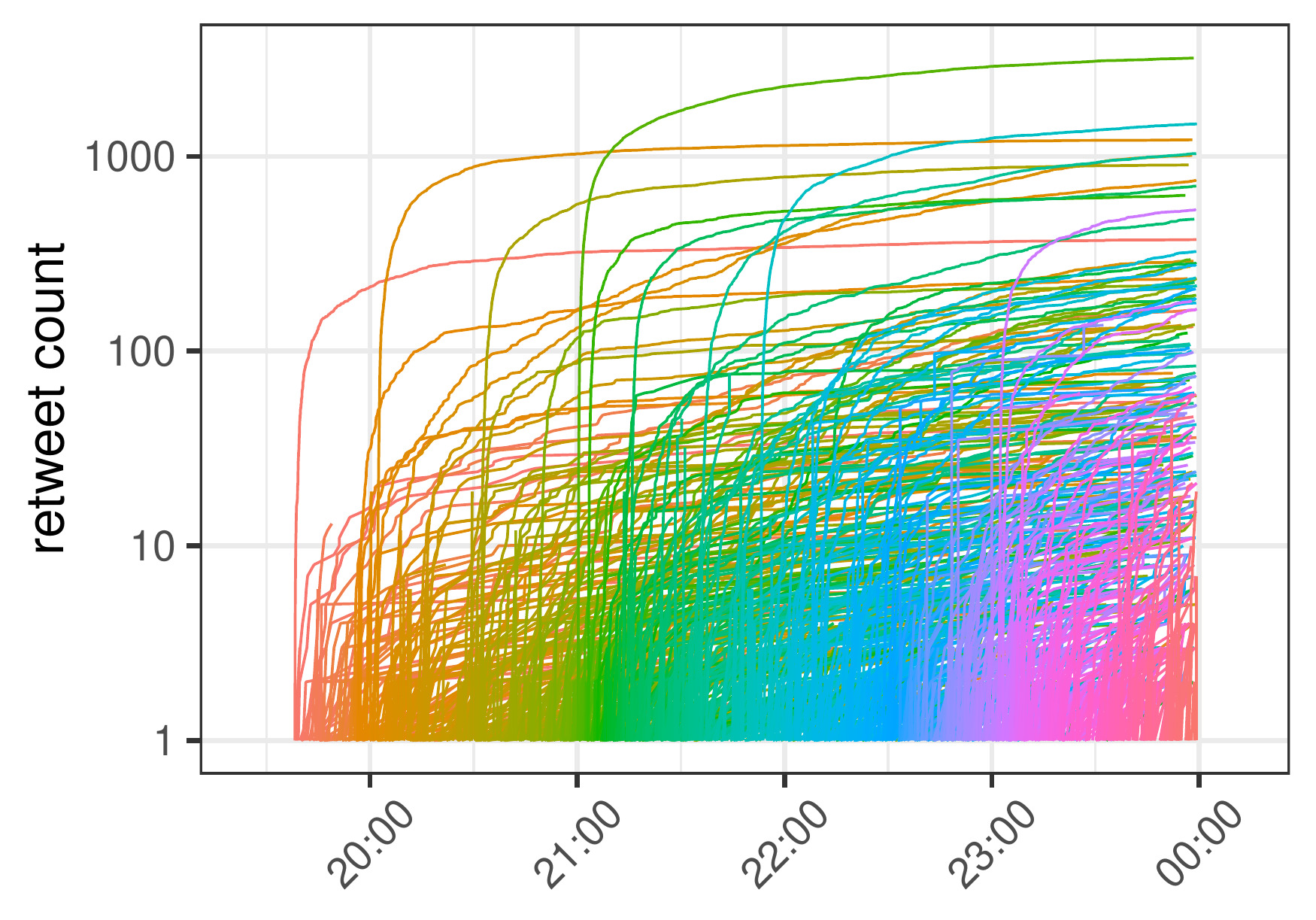} 

}

\caption[Cumulative retweet counts over time for \#thehandmaidstale (left) and \#gots7 (right) data]{Cumulative retweet counts over time for \#thehandmaidstale (left) and \#gots7 (right) data. Each trajectory is of a different colour and represents the growth of retweets of one individual original.}\label{fig.both_times}
\end{figure}

\end{knitrout}

\subsection{Hybrid process and the Duane plot}
Another way of visualising the temporality of the data is through the Duane plot \citep{duane64}. In order to do so we have to first introduce the hybrid process and the power law process. Consider a NHPP with intensity function
\begin{align}
  h(t)=\gamma t^{-\lambda}e^{-\theta t}, \label{eqn.intensity}
\end{align}
where $\gamma>0, \theta\geq0$ and $\lambda<1$, which is called the hybrid process hereafter. It is equivalent to the ``power law with exponential cutoff'' function by \cite{mmnb17}, but is different from the doubly stochastic processes introduced in Section \ref{sect.intro} as $h(t)$ is deterministic. The cumulative intensity is given by 
\begin{align}
  H(t):=\int_0^th(u)du=\left\{\begin{array}{ll}
  \gamma\Gamma(1-\lambda,\theta t)\theta^{\lambda-1}, & \theta>0,\\
  \gamma t^{1-\lambda}/(1-\lambda), & \theta=0,
  \end{array}\right.\nonumber
\end{align}
where $\Gamma(x,y)$ is the lower incomplete Gamma function such that $\underset{y\to\infty}{\lim}\Gamma(x,y)$ is equal to the Gamma function $\Gamma(x)$. Now assume a sequence of $n$ events is generated from the hybrid process in the time interval $[0,T]$, in which the $i$-th event occurs at time $t_i (i=1,2,\ldots,n)$, so that $0\leq t_1\leq t_2\leq\cdots\leq t_n\leq T.$ It is straightforward to write down the likelihood function:
\begin{align}
  &f(t_1,t_2,\ldots,t_n|\lambda,\theta,\gamma):=\exp\left[-H(T)\right]\times\prod_{i=1}^{n}h(t_i)\label{eqn.data_lik_general}\\
  &\qquad=\left\{\begin{array}{ll}
  \exp\left[-\gamma\Gamma(1-\lambda,\theta T)\theta^{\lambda-1}\right]\times\prod_{i=1}^{n}\gamma t_i^{-\lambda}e^{-\theta t_i}, & \theta>0,\\
  \exp\left[-\gamma T^{1-\lambda}/(1-\lambda)\right]\times\prod_{i=1}^n\gamma t_i^{-\lambda}, & \theta=0.
  \end{array}\right.\label{eqn.data_lik}
\end{align}
The derivation of the likelihood for general temporal point processes, which also applies to \eqref{eqn.data_lik_general}, can be found in, for example, \cite{dv03}, Proposition 7.2.III. When $\theta=0$, the hybrid process becomes the power law process \citep{blr92}. Each of the interarrival times follow a truncated Weibull distribution \citep{yrst05}. The power law process is different from a renewal process with power law distributed interarrival times, such as the event-modulated Poisson process termed by \cite{mr17}. 

Going back to the aforementioned sequence of events, if we want to check if it is generated by the power law process, we can fit the hybrid process and then formally test whether $\theta$ is 0 using some estimation approach. However, there is also a diagnostic plot for checking whether the power law process is appropriate with no model fitting required. Observe that the expected number of events at $t_i~(i=1,2,\ldots)$, denoted by $\displaystyle\E[N(t_i)]$, should be close to $i$, where $N(t)$ is the number of events in the interval $[0,t]$. Under the power law process, the former is given by $\displaystyle\E[N(t_i)]=H(t_i)=\frac{\gamma t_i^{1-\lambda}}{1-\lambda}$, and so, if it is equal to $i$, we have
\begin{align}
  \frac{\gamma t_i^{1-\lambda}}{1-\lambda} = i
  \quad&\Leftrightarrow\quad
  \frac{t_i}{i}=\frac{1-\lambda}{\gamma}t_i^{\lambda}\nonumber\\
  \quad&\Leftrightarrow\quad
  \log\left(\frac{t_i}{i}\right)=\log\left(\frac{1-\lambda}{\gamma}\right)+\lambda\log t_i. \label{eqn.data_duane}
\end{align}
This means that plotting $t_i/i$, which is termed mean time between failures (MTBF) in reliability theory, against $t_i$ on the log-log scale should give approximately a straight line with slope $\lambda$ and intercept $\log\left(1-\lambda\right)-\log(\gamma)$. This is called the Duane plot \citep{duane64}, which serves as a useful tool for diagnosing if the power law process describes the data well, and is similar to judging whether the Weibull distribution is useful according to the survival log-log plot in survival analysis.

\begin{knitrout}
\definecolor{shadecolor}{rgb}{0.969, 0.969, 0.969}\color{fgcolor}\begin{figure}[htbp!]

{\centering \includegraphics[width=0.49\linewidth]{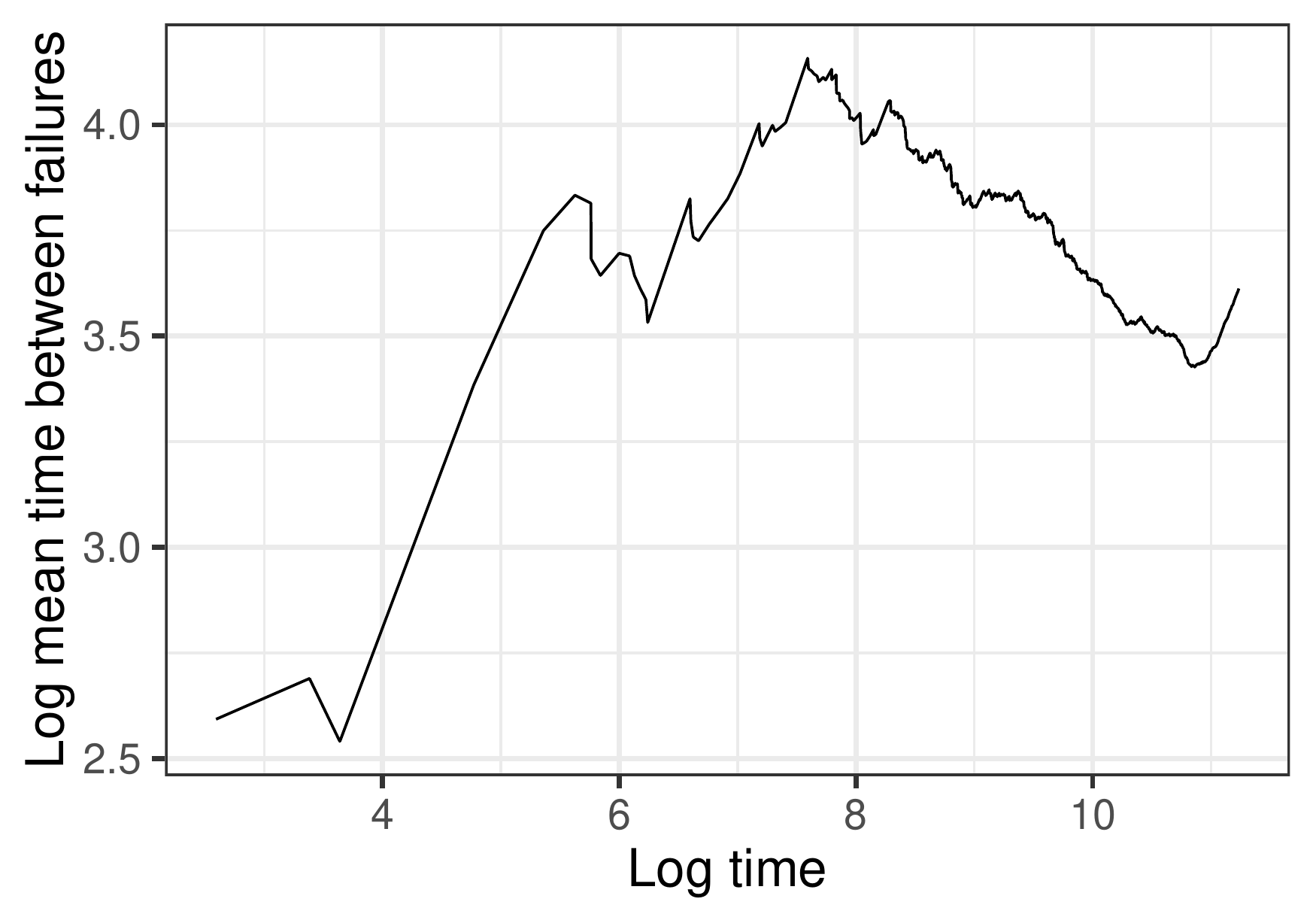} 
\includegraphics[width=0.49\linewidth]{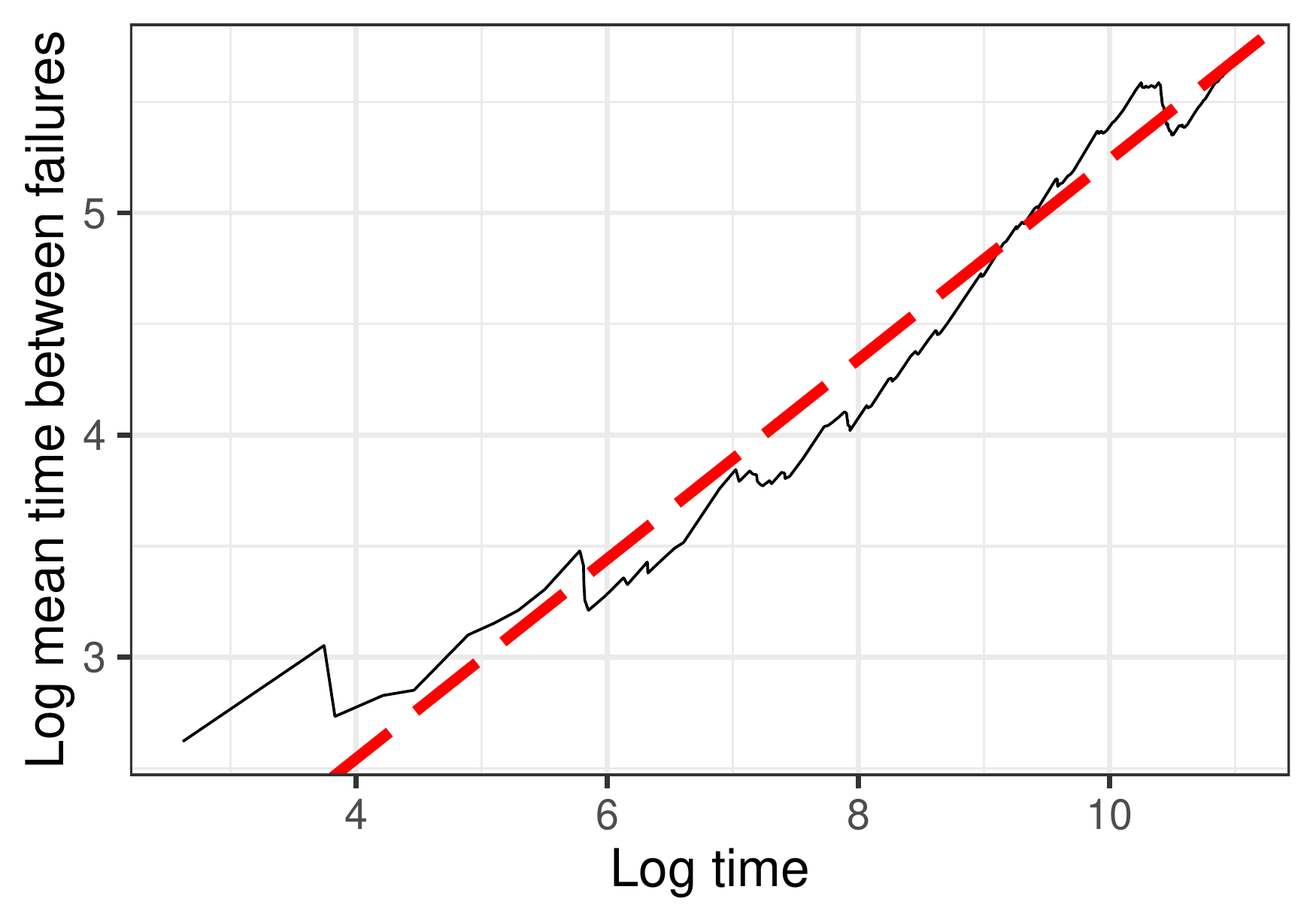} 

}

\caption{Duane plots of originals (left), where time is relative to the start of observation period, and retweets of the top original (right), where time is relative to when the original is tweeted, for \#thehandmaidstale data. Overlaid is the theoretical line (dashed) according to \eqref{eqn.data_duane} with $(\lambda,\gamma)=(\hat{\lambda},\hat{\gamma})$.}\label{fig.thehandmaidstale_duane}
\end{figure}

\end{knitrout}

The Duane plot for the 2043 originals of \#thehandmaidstale data is shown on the left of Figure \ref{fig.thehandmaidstale_duane}, where linearity is only observed at certain intervals. We also formally fit the power law process to the data, by maximising the (log-)likelihood in the second line of \eqref{eqn.data_lik} with respect to $\lambda$ and $\gamma$ simultaneously, yielding $(\hat{\lambda},\hat{\gamma})=(0.024, 0.034)$, where $\hat{\eta}$ denotes the maximum likelihood estimate (MLE) for any parameter $\eta$, and a maximised log-likelihood of -9422.9. Fitting the hybrid process instead gives $(\hat{\lambda},\hat{\theta},\hat{\gamma})=(-0.625, \ensuremath{3.152\times 10^{-5}}, \ensuremath{1.418\times 10^{-4}})$ and a maximised log-likelihood of -9336.4. While the difference in the maximised log-likelihood between the power law process and the hybrid process suggests that the former is inadequate, the latter does not necessarily describe the data well enough. 

On the right of Figure \ref{fig.thehandmaidstale_duane} is the Duane plot for the retweets of the top original (with 204 retweets) of \#thehandmaidstale data, which shows linearity over the whole observation period apart from a few small troughs and a seemingly increasing positive slope, suggesting that the power law process may be sufficient compared to the more general hybrid process. This is confirmed by fitting the latter to the data, which gives $(\hat{\lambda},\hat{\theta},\hat{\gamma})=(0.45, 0, 0.261)$, and no reduction in the maximised log-likelihood compared to the respective power law process fit. The dashed line overlaying the Duane plot represents \eqref{eqn.data_duane} with $(\lambda, \gamma)=(\hat{\lambda}, \hat{\gamma})$, and its proximity provides further support to the adequacy of the power law process.

\begin{knitrout}
\definecolor{shadecolor}{rgb}{0.969, 0.969, 0.969}\color{fgcolor}\begin{figure}[htbp!]

{\centering \includegraphics[width=0.49\linewidth]{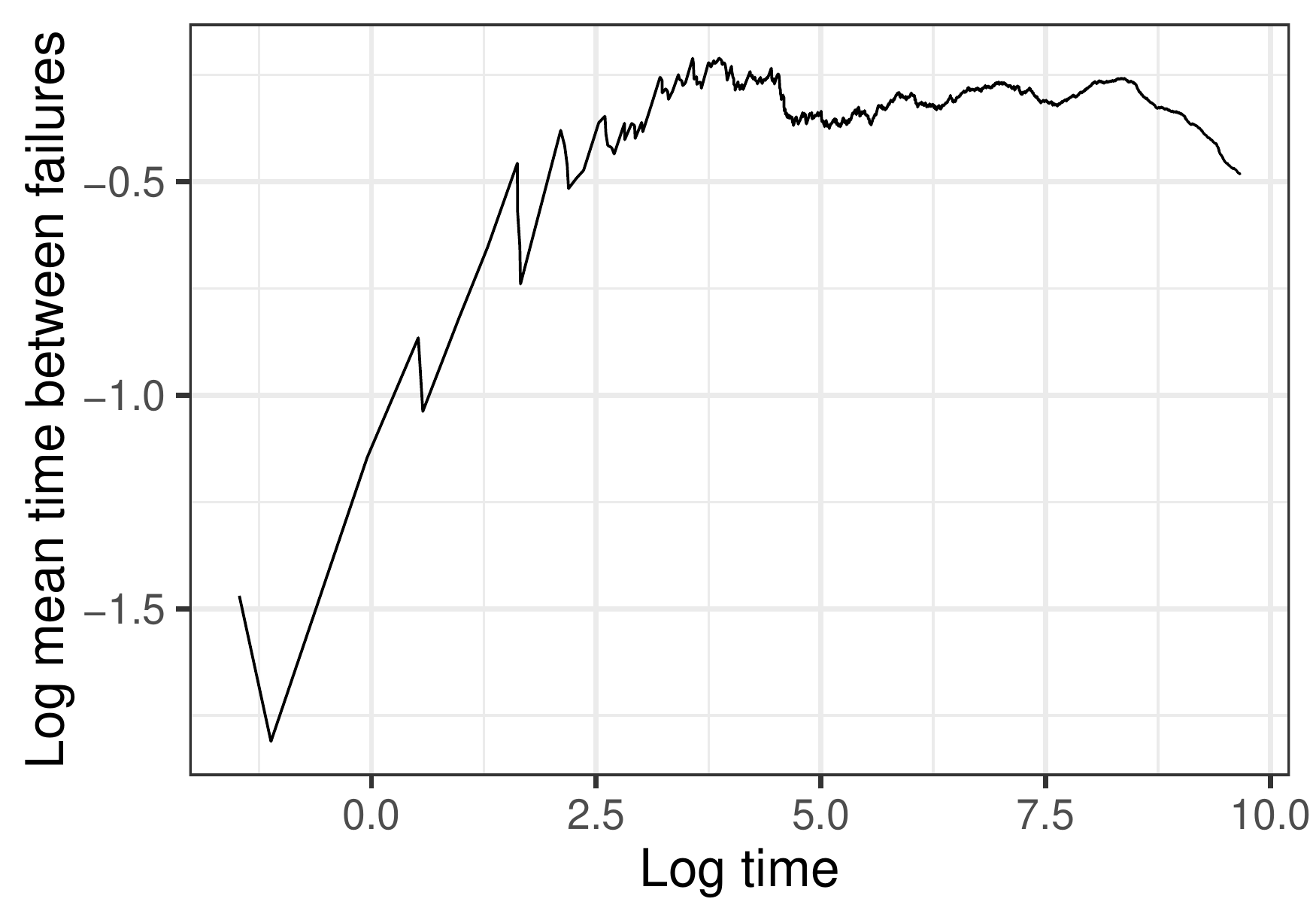} 
\includegraphics[width=0.49\linewidth]{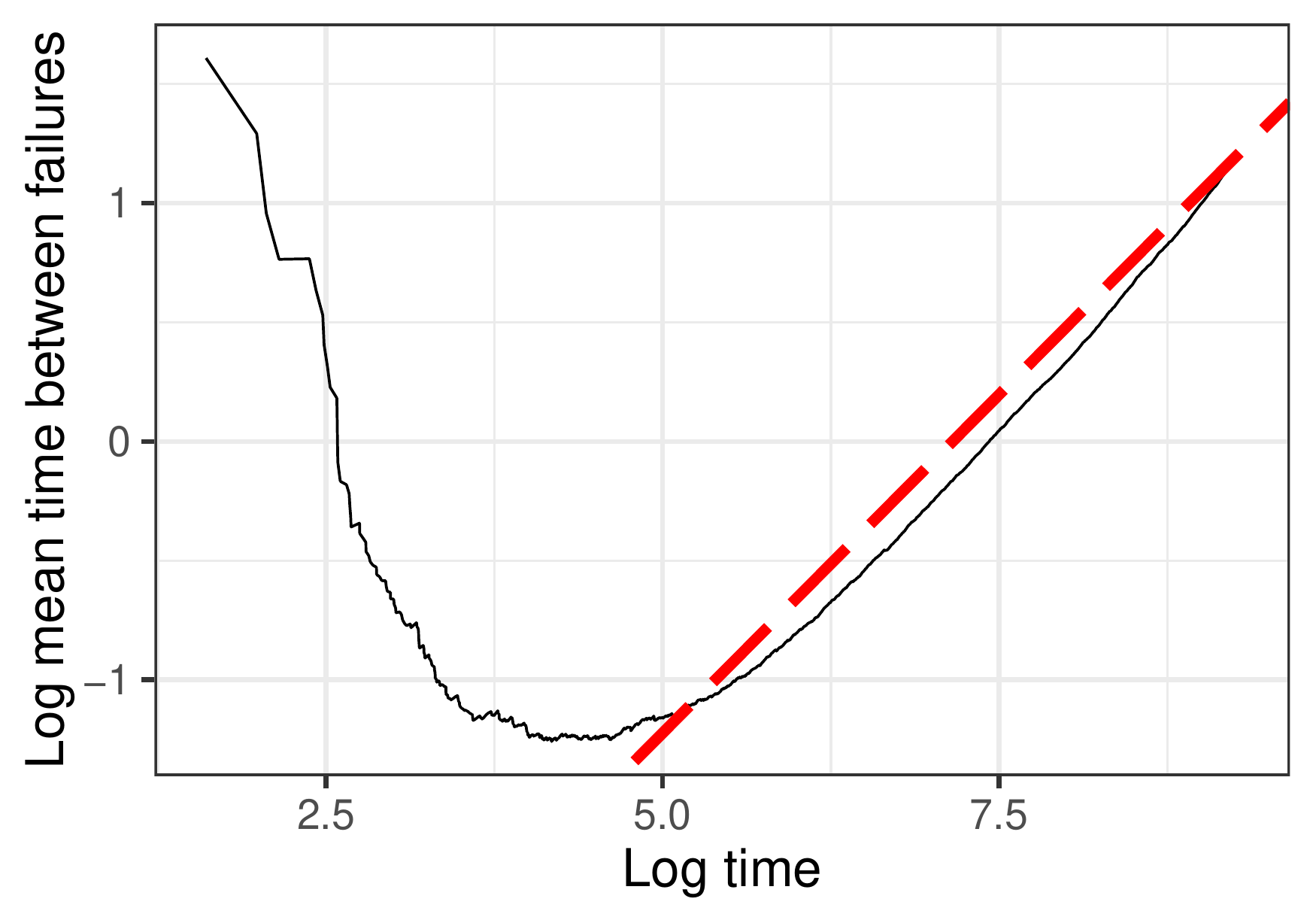} 

}

\caption{Duane plots of originals (left) and retweets of the top original (right) for the \#gots7 data, overlaid by the theoretical line (dashed) according to \eqref{eqn.data_duane}.}\label{fig.gots7_duane}
\end{figure}

\end{knitrout}

For the 25420 originals of \#gots7 data, fitting the power law process gives $(\hat{\lambda},\hat{\gamma})=(-0.131, 0.514)$, while the hybrid process does not improve the fit with the same point estimates and $\hat{\theta}=0$. For the 3204 retweets of the top original, fitting the power law process gives maximised log-likelihood -5545.8 and $(\hat{\lambda},\hat{\gamma})=(0.568, 25.104)$, which are used to obtain the theoretical line overlaid in the Duane plot in Figure \ref{fig.gots7_duane}. The slight concavity of the Duane plot in the overlapping interval indicates possible inadequacy of the power law process and potential improvement by the hybrid process, which is supported by fitting the latter to obtain maximised log-likelihood -5401.1 and $(\hat{\lambda},\hat{\theta},\hat{\gamma})=(0.408, \ensuremath{1.562\times 10^{-4}}, 12.932)$.

For each of the two hashtags considered, whether the power law process is adequate for the retweets of the top original should not be assumed to automatically apply to the retweets of every other original. For exploratory purposes, we overlay the Duane plots of retweets of the top 13 originals in Figure \ref{fig.gots7_duane_top_retweets}, all with over 300 retweets. That the slope is more similar across different Duane plots than the position is suggests that respective fits by the power law process (or the hybrid process) will give more similar estimates of $\lambda$ than of $\gamma$. In terms of actual modelling, we will generalise the hybrid process in the hierarchical model for the retweets introduced in the next section, and formally test whether $\theta$, which will be universal to all originals, is equal to 0 in Section \ref{sect.inf} through model selection.

\begin{knitrout}
\definecolor{shadecolor}{rgb}{0.969, 0.969, 0.969}\color{fgcolor}\begin{figure}[htbp!]

{\centering \includegraphics[width=0.5\linewidth]{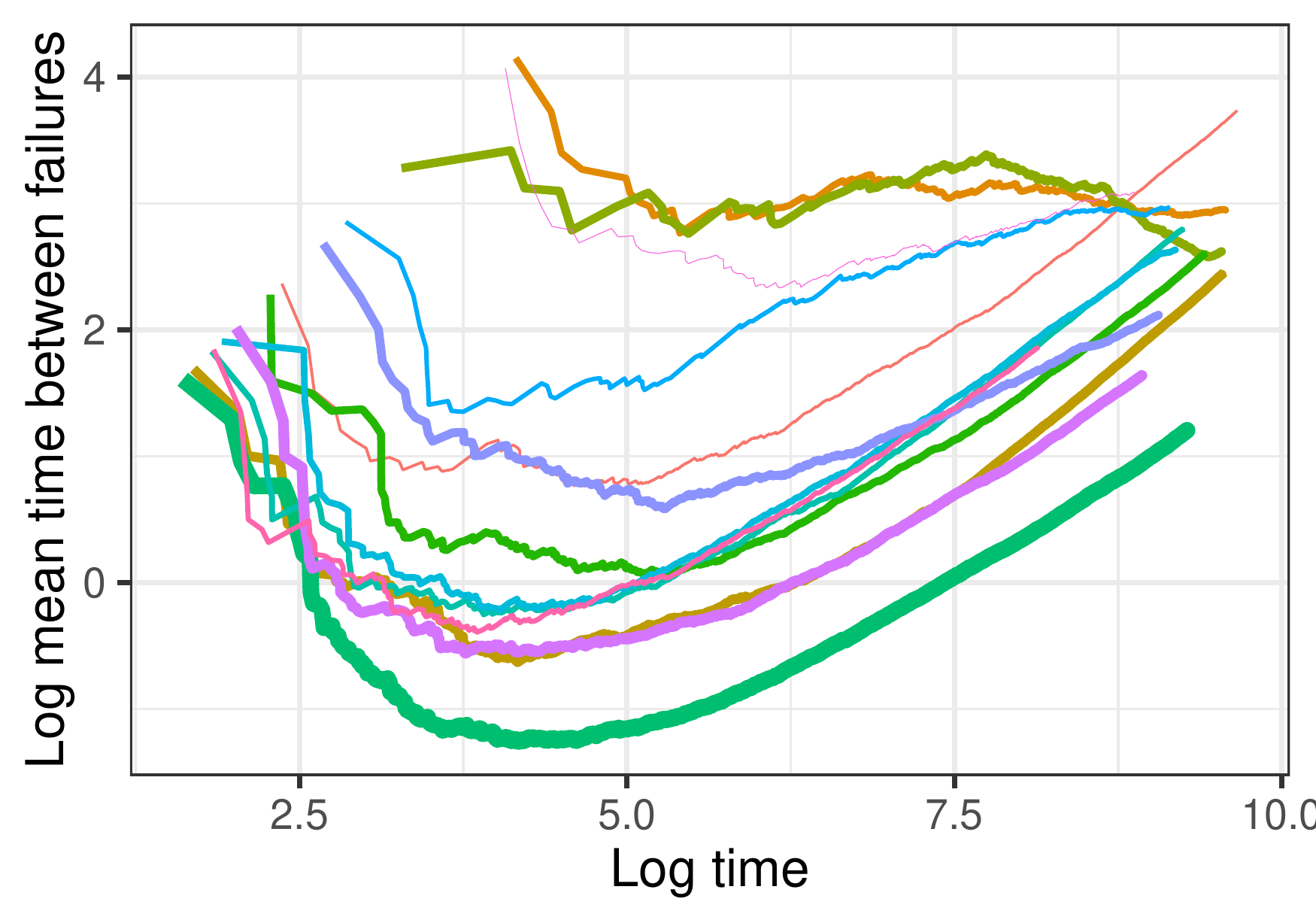} 

}

\caption[Duane plots of retweets of the top 13 originals for the \#gots7 data]{Duane plots of retweets of the top 13 originals for the \#gots7 data. The thicker the line is, the more retweets the original has.}\label{fig.gots7_duane_top_retweets}
\end{figure}

\end{knitrout}

The temporality of the originals and retweets aside, we are also interested in explaining the retweet count, which is the outcome of the process generating retweets, by the follower count, which is observed once the original is tweeted. We assume that the $i$-th original is tweeted at time $s_i$, when the author of which has $x_i^{*}\geq0$ followers. At time $T$, the end of the observation period, there are $m_i\geq0$ retweets observed for this original. Next, we define $m^{*}_i=\log(1+m_i)$ and $\displaystyle x_i=\log(1+x_i^{*})-\frac{1}{n}\sum_{k=1}^{n}\log(1+x_k^{*})$ to be the ``log'' retweet count and ``mean-centred'' follower count, respectively. While $x_i$ will be used as the covariate in the hierarchical modelling, the loose definition of the ``log'' retweet count is to ensure $m_i^{*}$ is finite, which will only be used for exploratory purposes in this section. The scatterplots of $m_i^{*}$ against $x_i$ for the retweets of the aforementioned data sets are shown in Figure \ref{fig.followers_retweet}, indicating that linear regression may be appropriate on these scales. The full model in Section \ref{sect.model} adheres to such relationships as it essentially models the retweet count $m_i$ according to a Poisson regression, in which the mean is proportional to the exponential of a linear predictor of the covariate $x_i$.

\begin{knitrout}
\definecolor{shadecolor}{rgb}{0.969, 0.969, 0.969}\color{fgcolor}\begin{figure}[htbp!]

{\centering \includegraphics[width=0.49\linewidth]{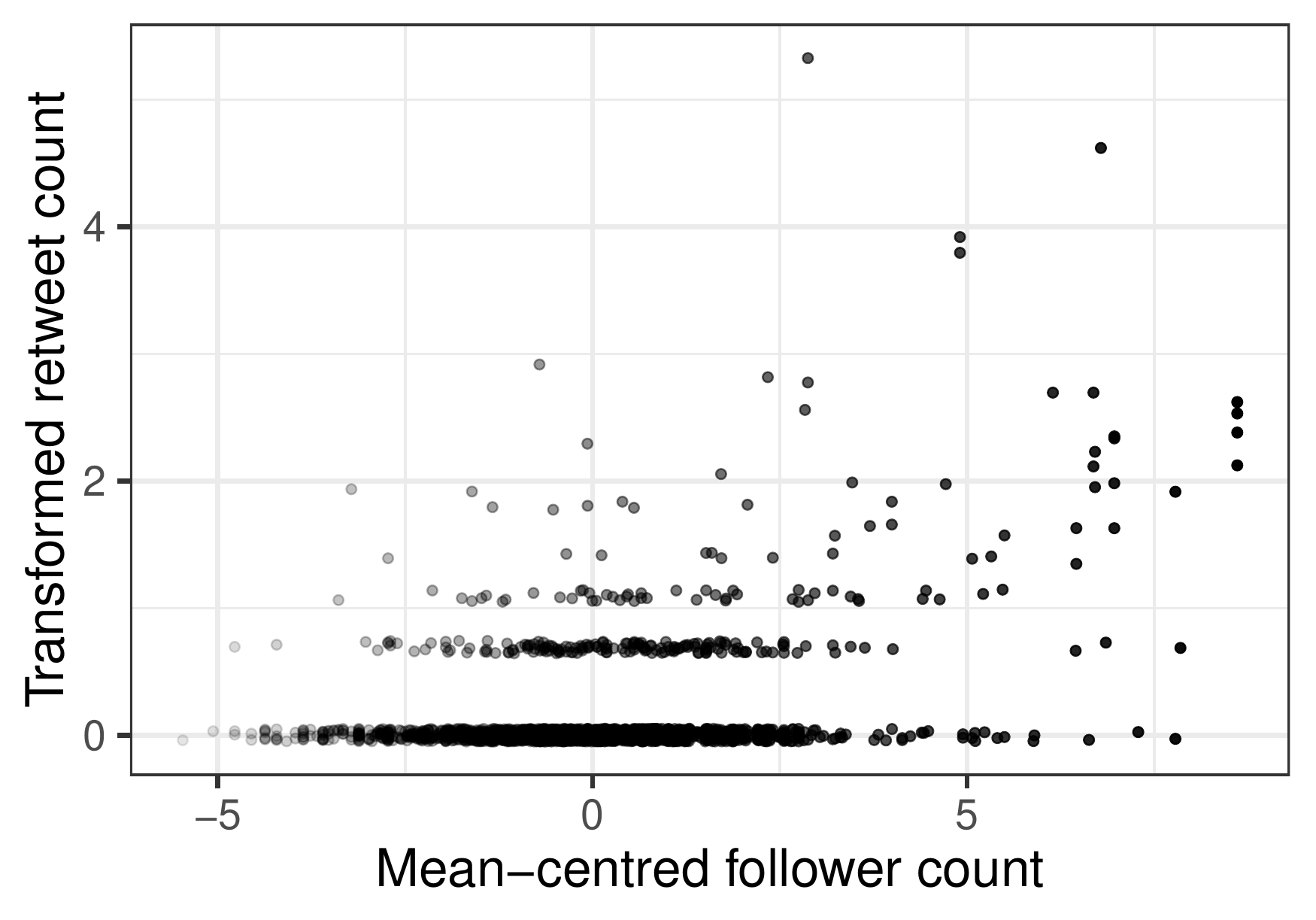} 
\includegraphics[width=0.49\linewidth]{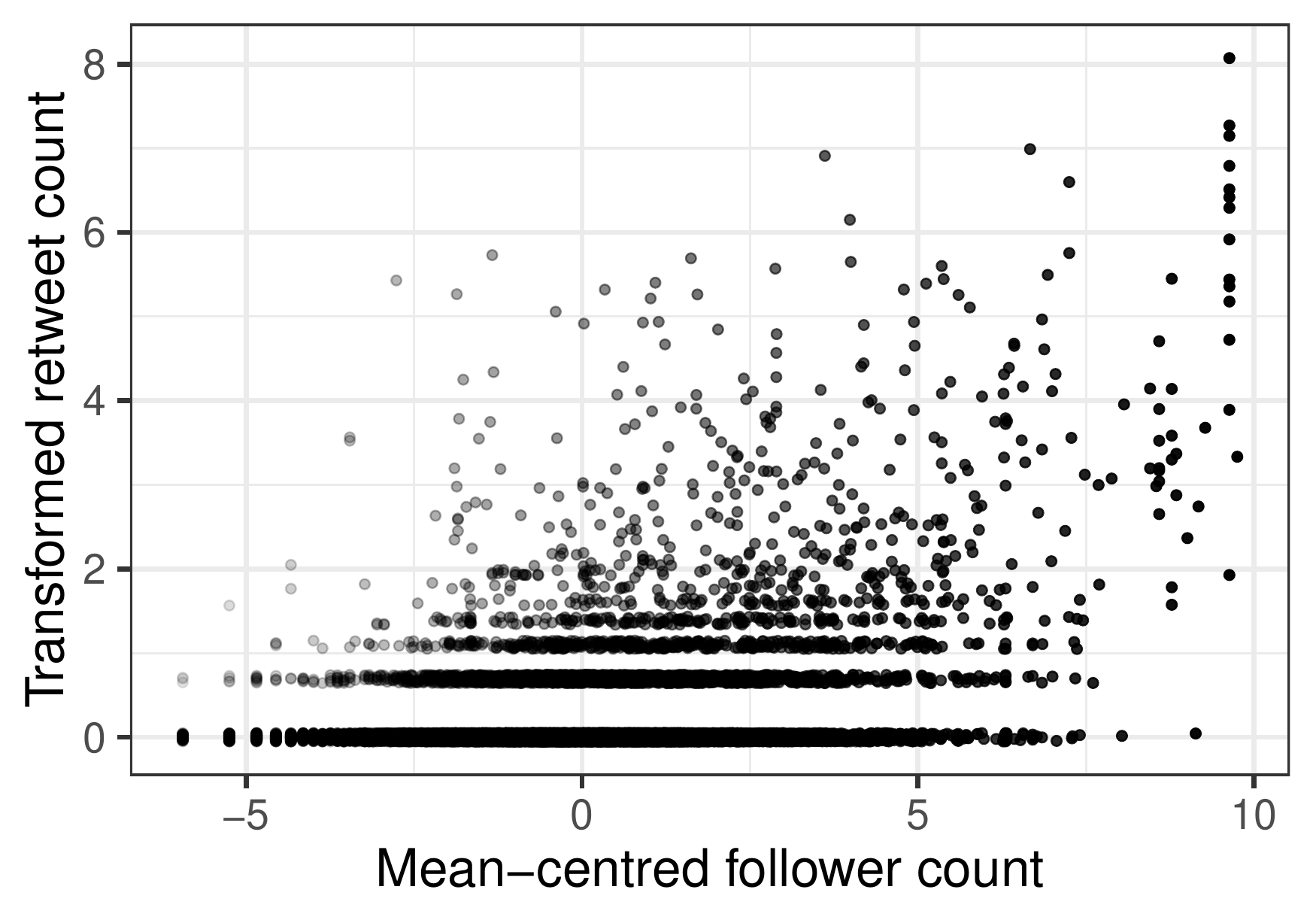} 

}

\caption[Log retweet count against mean-centred follower count for \#thehandmaidstale data (left) and \#gots7 data (right)]{Log retweet count against mean-centred follower count for \#thehandmaidstale data (left) and \#gots7 data (right).}\label{fig.followers_retweet}
\end{figure}

\end{knitrout}

\section{Hierarchical model and likelihood} \label{sect.model}
For convenience, the terminology and notation in Section \ref{sect.data} are retained, but no model is assumed for how the \textit{originals} are generated as it is not the concern of this section. Instead, modelled are the \textit{retweets} of the $i$-th original ($i=1,2,\ldots,n$) observed in $[s_i,T]$. The times of these $m_i$ retweets, denoted by $t_{i1},t_{i2},\ldots,t_{im_i}$ such that $s_i\leq t_{i1}\leq t_{i2}\leq\cdots\leq t_{im_i}\leq T$, are assumed to arise from a generalised hybrid process with intensity
\begin{align}
  h_i(t)=\phi~e^{\delta_i}(t-s_i+\psi)^{-\lambda}e^{-\theta(t-s_i)}\boldsymbol{1}_{\{t\geq s_i\}},\label{eqn.model_intensity}
\end{align}
where $\lambda<1$, $\phi>0$, $\psi\geq0$, $\theta\geq0$, and $\boldsymbol{1}_{\{A\}}$ is the indicator function of event $A$. The cumulative intensity at time $T$ follows directly:
\begin{align}
  &H_i(T)=\int_{0}^{T}h_i(t)dt=\int_{s_i}^{T}\phi~e^{\delta_i}(t-s_i+\psi)^{-\lambda}e^{-\theta(t-s_i)}dt\nonumber\\
  =~&\phi~e^{\delta_i}\times\left\{\begin{array}{ll}
  \left[(T-s_i+\psi)^{1-\lambda}-\psi^{1-\lambda}\right](1-\lambda)^{-1}, & \theta=0,\\
  \left[\Gamma\left(1-\lambda,\theta(T-s_i+\psi)\right)-\Gamma\left(1-\lambda,\theta\psi\right)\right]\theta^{\lambda-1}e^{\theta\psi}, & \theta>0.
  \end{array}\right. \label{eqn.model_cum_intensity}
\end{align}
Several modifications from \eqref{eqn.intensity} can be observed in \eqref{eqn.model_intensity}. First, the shift from $t$ to $t-s_i$ in $h_i(t)$ is due to the process of retweets taking place relative to the time original $i$ is tweeted. Second, while $\lambda$ and $\theta$ are universal to the process of each original, $\delta_i$ is dependent on $x_i$ and assumed to take the form
\begin{align}
  \delta_i&=\DELTA, \label{eqn.model_delta}\\
  \intertext{where}
  \epsilon_i~&\overset{\text{iid}}{\sim}~\text{N}(0,\tau^{-1}),\qquad i=1,2,\ldots,n. \label{eqn.model_e}
\end{align}
Third, there is no need for an intercept term in $\delta_i$ as it is embedded in $\phi$, and $\phi~e^{\delta_i}$ replaces $\gamma$ in \eqref{eqn.intensity} as the scale component for the retweet count of the $i$-th original. Finally, the inclusion of $\psi$, along with other generalisations, is for comparison with the ETAS model, proposed by \cite{ogata88} and mentioned in Section \ref{sect.intro}, later in this section.

Due to the nature of the NHPP, the retweet count $m_i$ for the $i$-th original at time $T$, follows the Poisson distribution with mean $H_i(T)=\phi~e^{\DELTA}\times$ (terms constant to $x_i$). Effectively and implicitly, a Poisson regression model is assumed for $m_i$ by $x_i$, even though the former is not directly modelled. The use of the NHPP enables the retweet times to be modelled while simultaneously explaining the retweet count by the ``mean-centred'' follower count.

If we sum all intensities of the retweet process of individual originals, we obtain the overall intensity of a point process of all retweets:
\begin{align}
  h(t) := \sum_{i:t>s_i}h_i(t) = \sum_{i:t>s_i}\phi~e^{\DELTA}\left(t-s_i+\psi\right)^{-\lambda}e^{-\theta(t-s_i)}. \label{eqn.all_intensity}
\end{align}
This seems similar to the \textit{conditional} intensity of the temporal ETAS model
\begin{align}
  h(t|\mathcal{H}_t) :=&~\mu+\sum_{k:t>t_k}\phi~e^{\beta(x_i-x_0)}\left(t-t_k+\psi\right)^{-\lambda},\label{eqn.etas_intensity}
\end{align}
where $\mathcal{H}_t$ is the history of all events up to time $t$. The parameterisation is slightly different from that usually seen in the literature \citep{ogata88, ca17, reinhart18}, for the sake of easier alignment. The apparent differences include the presence of $x_0$ in \eqref{eqn.etas_intensity}, which however is usually given in modelling earthquake data, the inclusion of an extra parameter $\kappa$ and a random effects term $\epsilon_i$, which will be justified in our application in Section \ref{sect.app}, and the additional exponential decay term over time $e^{-\theta(t-s_i)}$ suggested by \cite{mmnb17}. However, more important are the major differences in the underlying model structure. First, the ETAS model jointly models both the background events (originals in our case) and the triggered events (retweets) according to one point process, mainly because the nature of the events is not known prior to modelling, while in the proposed model concerned are the retweets \textit{given the originals}. Such difference can be seen in \eqref{eqn.all_intensity} that $\mu$ is absent, which under the ETAS model means the originals are modelled by a HPP. Second, while the times $s_i$ in \eqref{eqn.all_intensity} refer to the background events (originals) only, the times $t_k$ in \eqref{eqn.etas_intensity} refer to both type of events, hence the self-exciting nature of the process. This means under the ETAS model retweets can arise from other retweets, and different retweets essentially belong to a different and unobserved layer. On the other hand, under the proposed model, there is no self-excitation and only two layers of events exist, namely the originals from the background process, and the retweets that arise as offspring of the originals. Finally, under the ETAS model the layers, as well as the association between events of different layers, are unobserved and assumed by the model structure. Under the proposed model the relationships between the two completely known layers of events are observed and modelled accordingly.

Overall, our model deviates from the temporal ETAS model by removing the self-exciting nature of retweets. While it is possible to incorporate such structure, more information is required, such as the follower counts of the \textit{retweeters}, in order to compute a less tractable likelihood. See, for example, \cite{reinhart18} for the complexity of the required computations. Rather, we utilise the most valuable information, which is the correspondance between the originals and the retweets, to compute a completely tractable likelihood, which will be shown towards the end of this section, under a model that elegantly encompasses a collection of NHPPs of retweets.

It is useful to define a few vectors for the derivations in the rest of this section. We write
\begin{align}
  \boldsymbol{\eta}_0 &:= (\beta,\kappa,\lambda,\phi,\psi,\tau), \nonumber\\
  \boldsymbol{\eta}_1 &:= (\beta,\kappa,\lambda,\phi,\psi,\tau,\theta), \nonumber\\
  \boldsymbol{\epsilon} &:= (\epsilon_1,\epsilon_2,\ldots,\epsilon_n), \nonumber\\
  \boldsymbol{m} &:= (m_1,m_2,\ldots,m_n), \nonumber\\
  \boldsymbol{x} &:= (x_1,x_2,\ldots,x_n),\nonumber\\
  \boldsymbol{s} &:= (s_1,s_2,\ldots,s_n), \nonumber\\
  \boldsymbol{t}_i &:= (t_{i1},t_{i2},\ldots,t_{im_i}),\quad i=1,2,\ldots,n,\nonumber\\
  \intertext{and}
  \boldsymbol{t} &:= \{\boldsymbol{t}_1,\boldsymbol{t}_2,\ldots,\boldsymbol{t}_n\}.\nonumber
\end{align}
The two vectors $\boldsymbol{\eta}_0$ and $\boldsymbol{\eta}_1$ correspond to the generalised power law and hybrid processes, respectively. When there is no confusion, we simply write $\boldsymbol{\eta}$ to represent this vector of scalar parameters. While the four vectors $\boldsymbol{\epsilon}$, $\boldsymbol{m}$, $\boldsymbol{x}$ and $\boldsymbol{s}$ are all of length $n$, $\boldsymbol{\epsilon}$ is a vector of latent variables whereas the others are given as data/covariates. Finally, $\boldsymbol{t}_i (i=1,2,\ldots,n)$ is the vector of retweet times of the $i$-th original, while $\boldsymbol{t}$ is the collection of retweet times of \textit{all} originals. The length of $\boldsymbol{t}$, or equivalently the total number of retweets, is denoted by $m=\sum_{i=1}^{n}m_i$.

Before writing out the likelihood, we introduce a parameter $M$, which can take value 0 or 1, to represent model \textit{choice}. When $M=0$, the hierarchical model of the generalised power law process is the true model with parameter vector $\boldsymbol{\eta}_0$, in which $\theta$ is set to 0 and removed. When $M=1$, the hierarchical model of the generalised hybrid process with $\theta>0$ is the true model with parameter vector $\boldsymbol{\eta}_1$. By treating the nested models as two competing models, the problem of testing whether $\theta=0$ becomes a problem of model selection, which can be achieved by utilising the output of the inference algorithm outlined in Section \ref{sect.inf}. Our algorithm requires the likelihood for each model as a function of $\boldsymbol{\eta}_{M}$:
\begin{align}
  \begin{split}
    &f(\boldsymbol{m},\boldsymbol{t}|\boldsymbol{x},\boldsymbol{s},\boldsymbol{\epsilon},\boldsymbol{\eta}_0,M=0)\\
    &\qquad=\exp\left(-\frac{\phi}{1-\lambda}\sum_{i=1}^{n}e^{\DELTA}\left[(T-s_i+\psi)^{1-\lambda}-\psi^{1-\lambda}\right]\right)\\
    &\qquad\quad\times\phi^{m}\exp\left(\sum_{i=1}^{n}m_i\left[\DELTA\right]\right)\times\mathlarger\prod_{i:m_i>0}^{}~\mathlarger\prod_{j=1}^{m_i}\left(t_{ij}-s_i+\psi\right)^{-\lambda}, 
  \end{split}\label{eqn.model_lik_0}\\
  \begin{split}
    &f(\boldsymbol{m},\boldsymbol{t}|\boldsymbol{x},\boldsymbol{s},\boldsymbol{\epsilon},\boldsymbol{\eta}_1,M=1)\\
    &\qquad=\exp\left(-\phi~\theta^{\lambda-1}e^{\theta\psi}\sum_{i=1}^ne^{\DELTA}\left[\Gamma\left(1-\lambda,\theta(T-s_i+\psi)\right)-\Gamma\left(1-\lambda,\theta\psi\right)\right]\right)\\
    &\qquad\quad\times\phi^{m}\exp\left(\sum_{i=1}^{n}m_i\left[\DELTA\right]\right)\\
    &\qquad\quad\times\mathlarger\prod_{i:m_i>0}^{}~\mathlarger\prod_{j=1}^{m_i}\left(t_{ij}-s_i+\psi\right)^{-\lambda}\times\exp\left(-\theta\sum_{i:m_i>0}\sum_{j=1}^{m_i}\left(t_{ij}-s_{i}\right)\right).
  \end{split}\label{eqn.model_lik_1}
\end{align}
The derivations of \eqref{eqn.model_lik_0} and \eqref{eqn.model_lik_1} are detailed in Appendix A. Note that, even though $\tau$ is seen in neither \eqref{eqn.model_lik_0} nor \eqref{eqn.model_lik_1} because of independence between $\tau$ and the data conditional on $\boldsymbol{\epsilon}$, it is included in $\boldsymbol{\eta}_{M}$ for notational convenience.

\section{Inference and the Bayes factor} \label{sect.inf}

The presence of the latent variables $\boldsymbol{\epsilon}$ and the problem of model selection between $M=0$ and $M=1$ prompt us to consider Bayesian inference for the proposed hierarchical model. We first assign the following independent and vaguely informative priors:
\begin{align}
  \begin{array}{rcl}
    \beta&\sim&\text{N}\left(\mu_{\beta}=0,\tau_{\beta}^{-1}=\ensuremath{10^{4}}\right),\\
    \kappa&\sim&\text{N}\left(\mu_{\kappa}=0,\tau_{\kappa}^{-1}=\ensuremath{10^{4}}\right),\\
    (1-\lambda)&\sim&\text{Gamma}\left(a_{\lambda}=1,b_{\lambda}=0.001\right),\\
    \phi&\sim&\text{Gamma}\left(a_{\phi}=1,b_{\phi}=0.001\right),\\
    \psi&\sim&\text{Gamma}\left(a_{\psi}=1,b_{\psi}=0.001\right),\\
    \tau&\sim&\text{Gamma}\left(a_{\tau}=1,b_{\tau}=0.001\right),\\
    \theta&\sim&\text{Gamma}\left(a_{\theta}=1,b_{\theta}=0.001\right),\quad(\text{only for }M=1)
  \end{array}\label{eqn.inf_prior}
\end{align}
where $\tau_X^{-1}$ is the variance of a random variable $X\sim\text{N}\left(\mu_X,\tau_X^{-1}\right)$, and $a_Y/b_Y$ is the mean of a random variable $Y\sim\text{Gamma}\left(a_Y,b_Y\right)$. As the parameter space is not the same for $\boldsymbol{\eta}_{0}$ and $\boldsymbol{\eta}_{1}$, we denote $\boldsymbol{\eta}_{\backslash M}$ as the subset of $\boldsymbol{\eta}_M$ not in $\boldsymbol{\eta}_{1-M}$, which means $\boldsymbol{\eta}_{\backslash0}=\theta$ and $\boldsymbol{\eta}_{\backslash1}=\{\}$, the null set. Assuming conditional independence of $\boldsymbol{\eta}_{M}$ and $\boldsymbol{\eta}_{\backslash M}$ given $M$, the joint posterior of $\boldsymbol{\epsilon}$, $\boldsymbol{\eta}_{M}$, $\boldsymbol{\eta}_{\backslash M}$ and $M$ is
\begin{align}
  \begin{split}
    &\pi(\boldsymbol{\epsilon},\boldsymbol{\eta}_M,\boldsymbol{\eta}_{\backslash M},M|\boldsymbol{m},\boldsymbol{t},\boldsymbol{x},\boldsymbol{s})\propto\pi(\boldsymbol{m},\boldsymbol{t},\boldsymbol{\epsilon},\boldsymbol{\eta}_{M},\boldsymbol{\eta}_{\backslash M},M|\boldsymbol{x},\boldsymbol{s})\\
    &\qquad=f(\boldsymbol{m},\boldsymbol{t}|\boldsymbol{x},\boldsymbol{s},\boldsymbol{\epsilon},\boldsymbol{\eta}_{M},M)\times\pi(\boldsymbol{\epsilon}|\boldsymbol{\eta}_{M})\times\pi(\boldsymbol{\eta}_M|M)\times\pi(\boldsymbol{\eta}_{\backslash M}|M)\times\pi(M),
  \end{split}\label{eqn.inf_joint_post}
\end{align}
where $\pi(\boldsymbol{\epsilon}|\boldsymbol{\eta}_M)$ can be simplified to $\pi(\boldsymbol{\epsilon}|\tau)$ and is given by \eqref{eqn.model_e}, while the last component $\pi(M)$ is the prior model probability. The ``true'' prior of $\boldsymbol{\eta}_{M}$ under $M$ is given by
\begin{align}
  \pi(\boldsymbol{\eta}_{M}|M)&=\left\{
  \begin{array}{ll}
    \pi_{\beta}(\beta)\pi_{\kappa}(\kappa)\pi_{\lambda}(\lambda)\pi_{\psi}(\psi)\pi_{\phi}(\phi)\pi_{\tau}(\tau), & M=0,\\
    \pi_{\beta}(\beta)\pi_{\kappa}(\kappa)\pi_{\lambda}(\lambda)\pi_{\psi}(\psi)\pi_{\phi}(\phi)\pi_{\tau}(\tau)\pi_{\theta}(\theta|M=1), & M=1,
  \end{array}
  \right.\nonumber
\end{align}
where the components are given by \eqref{eqn.inf_prior}. For $M=1$, the pseudoprior $\pi(\boldsymbol{\eta}_{\backslash M}|M)$ vanishes as $\boldsymbol{\eta}_{\backslash 1}=\{\}$, while for $M=0$, the pseudoprior can be written as $\pi_{\theta}(\theta|M=0)$ equivalently. We proceed to draw samples of $(\boldsymbol{\epsilon},\boldsymbol{\eta}_M,\boldsymbol{\eta}_{\backslash M},M)$ using Markov chain Monte Carlo (MCMC), in which model selection is facilitated by the modified version \citep{dfn02} of Gibbs variable selection \citep{cc95}. The MCMC algorithm is outlined as follows:
\begin{enumerate}
  \item The current values in the chain are $\boldsymbol{\epsilon}$, $\boldsymbol{\eta}_M$, $\boldsymbol{\eta}_{\backslash M}$ and $M$.
  \item \label{item.mcmc_cond_post} Draw $\boldsymbol{\eta}_M$ from its conditional posterior, with density proportional to \\$f(\boldsymbol{m},\boldsymbol{t}|\boldsymbol{x},\boldsymbol{s},\boldsymbol{\epsilon},\boldsymbol{\eta}_M,M)\times\pi(\boldsymbol{\epsilon}|\tau)\times\pi(\boldsymbol{\eta}_M|M)$, by a fairly standard component-wise Metropolis-within-Gibbs (MWG) algorithm, the details of which are given in Appendix B. Denote the value by $\boldsymbol{\eta}_M^{'}$.
  \item \label{item.mcmc_e} Draw $\boldsymbol{\epsilon}$ from its conditional posterior, with density proportional to \\$f(\boldsymbol{m},\boldsymbol{t}|\boldsymbol{x},\boldsymbol{s},\boldsymbol{\epsilon},\boldsymbol{\eta}_M,M)\times\pi(\boldsymbol{\epsilon}|\tau)$, by the same MWG alogithm in Appendix B. Denote the value by $\boldsymbol{\epsilon}^{'}$.
  \item \label{item.mcmc_pseudoprior} If $M=0$, draw $\theta$ from its pseudoprior $\pi_{\theta}(\theta|M=0)$. Denote the value by $\theta^{'}$, and write $\boldsymbol{\eta}_{1}^{'}=(\boldsymbol{\eta}_{0}^{'},\theta^{'})$. If $M=1$, write $\boldsymbol{\eta}_0^{'}=\boldsymbol{\eta}_{1,-\theta}^{'}$, that is, the proposed value of $\boldsymbol{\eta}_1$ with that of $\theta$ dropped, so that $\boldsymbol{\eta}_{1}^{'}=(\boldsymbol{\eta}_{0}^{'},\theta^{'})$ still holds.
  \item \label{item.mcmc_m} Draw $M$ from $\pi(M|\boldsymbol{m},\boldsymbol{t},\boldsymbol{x},\boldsymbol{s},\boldsymbol{\epsilon},\boldsymbol{\eta}_M,\boldsymbol{\eta}_{\backslash M})$, its conditional posterior distribution. Essentially, set $M$ to 0 and 1 with probabilities $\displaystyle\frac{A_0}{A_0+A_1}$ and $\displaystyle\frac{A_1}{A_0+A_1}$, respectively, where, using \eqref{eqn.inf_joint_post},
\end{enumerate}
\begin{align}
  A_0&=f(\boldsymbol{m},\boldsymbol{t}|\boldsymbol{x},\boldsymbol{s},\boldsymbol{\epsilon}^{'},\boldsymbol{\eta}_0^{'},M=0)~\pi_{\theta}(\theta^{'}|M=0)~\pi(M=0),\nonumber\\
  A_1&=f(\boldsymbol{m},\boldsymbol{t}|\boldsymbol{x},\boldsymbol{s},\boldsymbol{\epsilon}^{'},\boldsymbol{\eta}_1^{'},M=1)~\pi_{\theta}(\theta^{'}|M=1)~\pi(M=1).\nonumber
\end{align}
\begin{enumerate}[resume]
  \item Denote the drawn value in step \ref{item.mcmc_m} by $M^{'}$. The current values are now $\boldsymbol{\epsilon}^{'}$, $\boldsymbol{\eta}_M^{'}$, $\boldsymbol{\eta}_{\backslash M}^{'}$ and $M^{'}$.
\end{enumerate}

The pseudoprior $\pi_{\theta}(\theta|M=0)$ is chosen to be close to the marginal posterior of $\theta$ under the competing model, denoted by $\pi_{\theta}(\theta|M=1,\boldsymbol{m},\boldsymbol{t},\boldsymbol{x},\boldsymbol{s})$, for the sake of optimisation \citep{dfn02, cc95}. It can be informed by a pilot run of the MWG algorithm in Appendix B for model $1$ individually. As the priors for the overlapping parameters are the same for both models, only the pseudoprior $\pi_{\theta}(\theta|M=0)$ and the prior $\pi_{\theta}(\theta|M=1)$ are involved in step \ref{item.mcmc_m}.

The draws of $\boldsymbol{\eta}_0$ in the above algorithm where $M=0$ marginally represent an approximate sample from $\pi(\boldsymbol{\eta}_0|M=0,\boldsymbol{m},\boldsymbol{t})$, that is, its posterior distribution \textit{under that model 0 is true}; likewise for $\boldsymbol{\eta}_1$. What is more important, however, is the empirical proportion of $M$, denoted by $\hat{\pi}(M|\boldsymbol{m},\boldsymbol{t})$, as it approximates the posterior probability that model $M$ is true. Finally, the Bayes factor is the ratio of the posterior odds to the prior odds:
\begin{align}
  B_{10}=\frac
  {\hat{\pi}(M=1|\boldsymbol{m},\boldsymbol{t},\boldsymbol{x},\boldsymbol{s})}
  {\hat{\pi}(M=0|\boldsymbol{m},\boldsymbol{t},\boldsymbol{x},\boldsymbol{s})}
  \left/\frac
  {\pi(M=1)}
  {\pi(M=0)}
  \right..
  \label{eqn.inf_bayes_factor}
\end{align}

An alternative to Gibbs variable selection (GVS) for model selection is reversible jump Markov chain Monte Carlo (RJMCMC) \citep{green95}, which should theoretically give the same posterior probabilities for the model choice. It will be used to verify with the results of GVS in the application, and the details of its algorithm are given in Appendix C.

\section{Model diagnostics} \label{sect.gof}

In this section, we augment the function arguments of the cumulative intensity in \eqref{eqn.model_cum_intensity} by writing $H_i(T;x_i,s_i,\epsilon_i,\boldsymbol{\eta}):=H_i(T)$, where $x_i$, $s_i$, $\epsilon_i$ and $\boldsymbol{\eta}$ are included whenever necessary. Under the proposed model, the random variable of the retweet count $m_i$ at time $T$ is Poisson distributed with mean (and variance) $H_i(T;x_i,s_i,\epsilon_i,\boldsymbol{\eta})$ which, according to \eqref{eqn.model_cum_intensity}, is
\begin{align}
  &\E_{m_i}\left[m_i|x_i,s_i,\epsilon_i,\boldsymbol{\eta},T\right]:=H_i(T;x_i,s_i,\epsilon_i,\boldsymbol{\eta})=H_i(T;x_i,s_i,0,\boldsymbol{\eta})\times e^{\epsilon_i}.\label{eqn.m_cond_mean}
\end{align}
The second equality can be seen by substituting \eqref{eqn.model_delta} into \eqref{eqn.model_cum_intensity}. As $\epsilon_i$ is N$(0,\tau^{-1})$ distributed \textit{apriori}, $e^{\epsilon_i}$ is log-normally distributed with mean $e^{0.5/\tau}$ and variance $e^{2/\tau}-e^{1/\tau}$. This enables us to obtain the expectation and variance of the expected retweet count in \eqref{eqn.m_cond_mean} \textit{with respect to $\epsilon_i$}:
\begin{align}
  &\E_{\epsilon_i}\left[\E_{m_i}\left[m_i\left|x_i,s_i,\epsilon_i,\boldsymbol{\eta},T\right.\right]\right]=\E_{\epsilon_i}\left[H(T;x_i,s_i,0,\boldsymbol{\eta})\times e^{\epsilon_i}\right]\nonumber\\
  =~&H_i(T;x_i,s_i,0,\boldsymbol{\eta})\times e^{0.5/\tau},\nonumber\\
  &\Var_{\epsilon_i}\left[\E_{m_i}\left[m_i\left|x_i,s_i,\epsilon_i,\boldsymbol{\eta},T\right.\right]\right]=\Var_{\epsilon_i}\left[H(T;x_i,s_i,0,\boldsymbol{\eta})\times e^{\epsilon_i}\right]\nonumber\\
  =~&\left(H_i(T;x_i,s_i,0,\boldsymbol{\eta})\right)^2\times\left(e^{2/\tau}-e^{1/\tau}\right).\nonumber
\end{align}
These two quantities can be used to obtain the $\chi^{2}$ discrepancy, which is a goodness-of-fit measure advocated by \cite*{gms96}:
\begin{align}
  \chi^2&=\sum_{i=1}^{n}\frac{\left(m_i-\E_{\epsilon_i}\left[\E_{m_i}\left[m_i\left|x_i,s_i,\epsilon_i,\boldsymbol{\eta},T\right.\right]\right]\right)^2}{\Var_{\epsilon_i}\left[\E_{m_i}\left[m_i\left|x_i,s_i,\epsilon_i,\boldsymbol{\eta},T\right.\right]\right]}\nonumber\\
  &=\left(e^{2/\tau}-e^{1/\tau}\right)^{-1}\sum_{i=1}^{n}\left(\frac{m_i}{H_i\left(T;x_i,s_i,0,\boldsymbol{\eta}\right)}-e^{0.5/\tau}\right)^2.\label{eqn.chisq}
\end{align}
This $\chi^2$ discrepancy is directly computable using the samples of $\boldsymbol{\eta}$ and $\boldsymbol{\epsilon}$ from the MCMC outlined in Section \ref{sect.inf}. At each iteration, using the observed retweet counts as $\boldsymbol{m}=(m_1,m_2,\ldots,m_n)$ and the current values of $\boldsymbol{\eta}$, the actual discrepancy, denoted by $\chi^2_{\text{act}}$, can be obtained. On the other hand, using the current values of $\boldsymbol{\eta}$ and $\epsilon_i$, we can simulate the retweet count, for the $i$-th original, from the Poisson distribution with mean $\E_{m_i}\left[m_i|x_i,s_i,\epsilon_i,\boldsymbol{\eta},T\right]=H_i(T;x_i,s_i,\epsilon_i,\boldsymbol{\eta})$. Plugging the whole set of simulated retweet counts as $\boldsymbol{m}$ and the current values of $\boldsymbol{\eta}$ into \eqref{eqn.chisq} yields the simulated discrepancy, denoted by $\chi^2_{\text{sim}}$. Finally, comparing the two sets of discrepancies will help us determine if there are any inadequecies of the model fit. This can be achieved by plotting $\chi^2_{\text{sim}}$ against $\chi^2_{\text{act}}$ and computing the associated posterior predictive $p$-value, which is the empirical proportion of $\chi^2_{\text{sim}}>\chi^2_{\text{act}}$. Both of these will be presented in Section \ref{sect.app}. For details of diagnostics for Bayesian hierarchical models in general, please see, for example, \cite{gms96} and Section 2.2.2 of \cite{cw11}.

\section{Application} \label{sect.app}

\begin{knitrout}
\definecolor{shadecolor}{rgb}{0.969, 0.969, 0.969}\color{fgcolor}\begin{figure}[htbp!]

{\centering \includegraphics[width=0.4\linewidth]{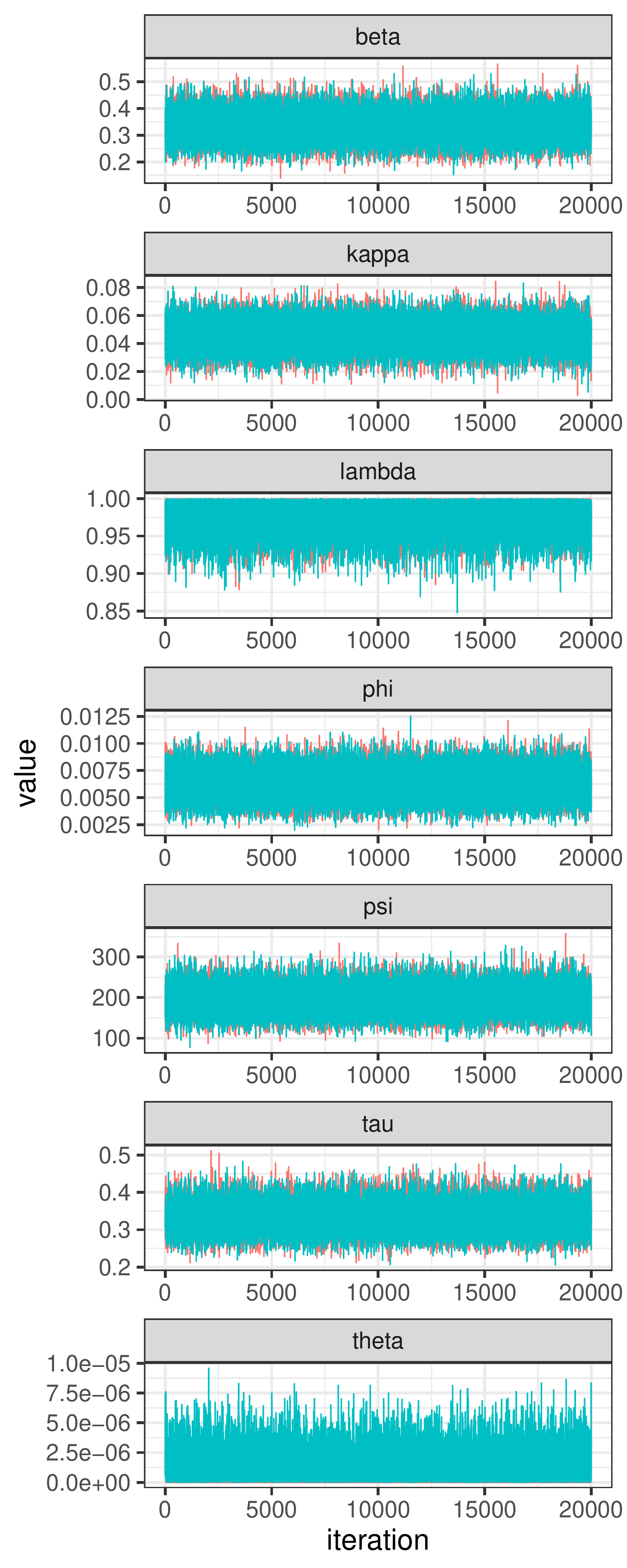} 
\includegraphics[width=0.4\linewidth]{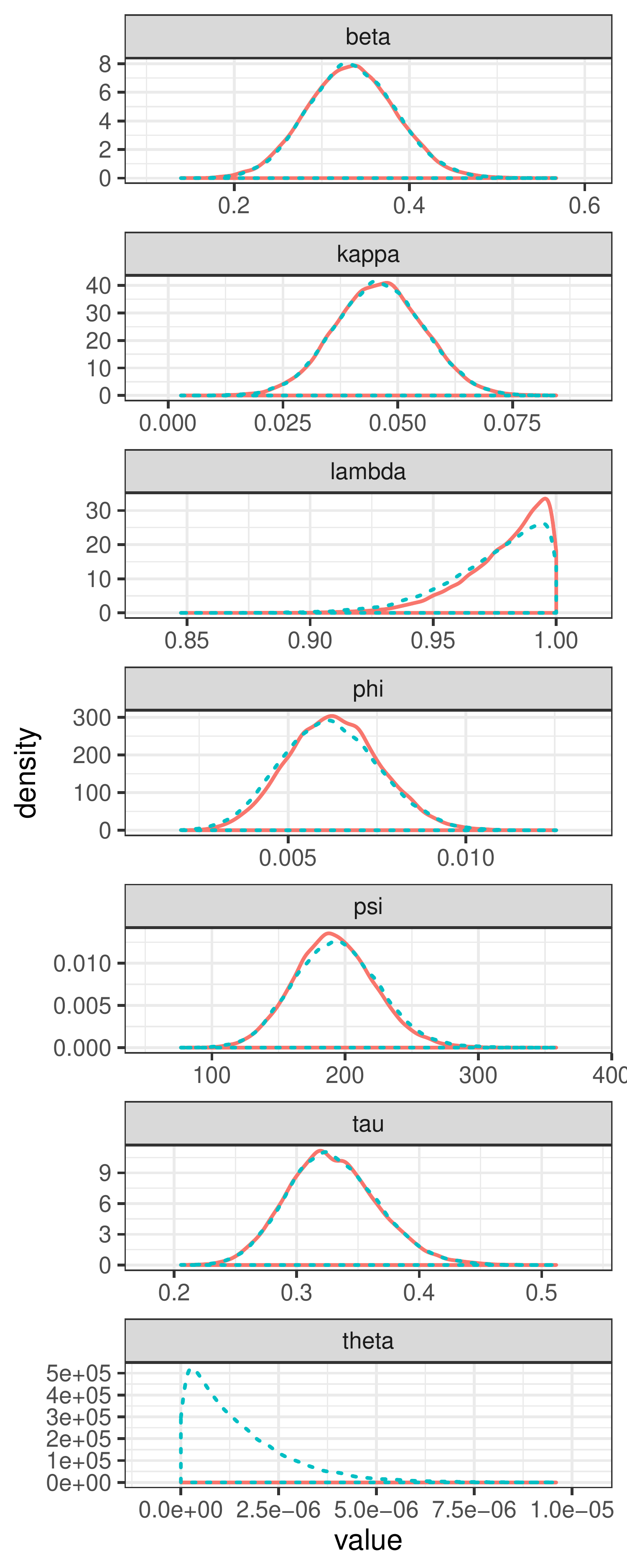} 

}

\caption[Traceplots (left) and posterior densities (right) of the parameters of models 0 (salmon, solid) and 1 (turquoise, dashed on the right) fitted to \#thehandmaidstale data]{Traceplots (left) and posterior densities (right) of the parameters of models 0 (salmon, solid) and 1 (turquoise, dashed on the right) fitted to \#thehandmaidstale data.}\label{fig.thehandmaidstale_mcmc_plot}
\end{figure}

\end{knitrout}

Both the model-specific algorithms in Appendix B and the model selection algorithm in \ref{sect.inf} are applied to the two data sets with different hashtags. \added{All MCMC runs were performed on a Linux machine with Intel Core i5-4690S Processor (3.2GHz). }For the \#thehandmaidstale data, each of the three algorithms is applied to the times of creation of the 2043 originals, 265 of which have been retweeted at least once, and of their associated retweets, to obtain a single chain of 20000 iterations, upon thinning of 2000, after discarding the first 1000000 as burn-in. The individual model fits are reported in the form of traceplots and posterior densities of the parameters in Figure \ref{fig.thehandmaidstale_mcmc_plot}\added{, and the computation times are reported in Table \ref{tab.mcmc_summaries}}. While the inclusion of $\theta$ in model 1 makes a substantial difference in terms of the posterior densities of the other parameters, what is more important is how the evidence of each model weighs against each other. In the model selection algorithm, the prior probabilities $\pi(M=0)$ and $\pi(M=1)$ are chosen artifically to be $\ensuremath{10^{-9}}$ and $1-\ensuremath{10^{-9}}$, respectively. They are chosen this way not to represent our prior belief in the models, but to ensure sufficient mixing between the two states of $M$ in the chain. Model 0 is selected for 7584 times out of 20000 iterations, meaning that the Bayes factor $B_{10}$ in \eqref{eqn.inf_bayes_factor} is estimated to be $\frac{12416}{7584}\left/\frac{1.0-\ensuremath{10^{-9}}}{\ensuremath{10^{-9}}}\right.=\ensuremath{1.637\times 10^{-9}}$. The RJMCMC algorithm gives a similar estimate of $B_{10}=\frac{12402}{7598}\left/\frac{1.0-\ensuremath{10^{-9}}}{\ensuremath{10^{-9}}}\right.=\ensuremath{1.632\times 10^{-9}}$. So, for the \#thehandmaidstale data set, which consists of tweets for over 21 hours, the generalised power law process hierarchical model is more appropriate. Such findings are different from the exponential cutoff phenomenon shown by \cite{mmnb17} for tweets collected over a similar duration of 24 hours.

\begin{table}[htbp!]
  \centering
  \renewcommand{\arraystretch}{0.75}
  \begin{tabular}{|l|c|c|}
    \hline
    & \#thehandmaidstale & \#gots7 \\ \hline\hline
    Originals ($n$) & 2043 & 25420 \\ \hline
    Total retweets ($m$) & 971 & 29751\\ \hline\hline
    Burn-in & 1000000 & 1000000 \\ \hline
    Thinning & 2000 & 1000 \\ \hline
    Length of thinned chain & 20000 & 20000 \\ \hline
    Total number of iterations & 41000000 & 21000000 \\ \hline\hline
    Time for model 0 & 28.2 & 221.5 \\ \hline
    Time for model 1 & 69 & 446.8 \\ \hline
    Time for GVS & 67.6 & 464.8 \\ \hline
    Time for RJMCMC & 34.7 & 239.9 \\ \hline
  \end{tabular}
  \caption{\added{Summaries of chains and computation times (in hours) of the MCMC algorithms for the two data sets. The numbers of originals $n$ and total retweets $m$ are the same as in Table \ref{tab.summaries}.}}
  \label{tab.mcmc_summaries}
\end{table}

The \#gots7 data set consists of 25420 originals, 3145 of which have been retweeted once, and 29751 retweets. \replaced{For}{To exploit parallelisation due to the much greater computational burden, for} each of the two model-specific algorithms and the model selection algorithm, \replaced{a chain of 20000 iterations is obtained}{1 chains are obtained, each of which contains 20000 iterations}, upon thinning of 1000, after discarding the first 1000000 as burn-in. The traceplots and the posterior densities for the parameters are plotted in Figure \ref{fig.gots7_mcmc_plot}. The proximity of the posterior densities for all parameters other than $\theta$ is similar to that for the \#thehandmaidstale data, suggesting that the inclusion of $\theta$ does not improve fit much. This is supported by the model selection results via GVS. With the prior probabilities $\pi(M=0)$ and $\pi(M=1)$ chosen to be $\ensuremath{10^{-10}}$ and $1-\ensuremath{10^{-10}}$, respectively, model 0 is selected for 7344 times out of 20000 iterations in total, meaning that the Bayes factor is estimated to be $B_{10}=\frac{12656}{7344}\left/\left(\frac{1-\ensuremath{10^{-10}}}{\ensuremath{10^{-10}}}\right)\right.=\ensuremath{1.723\times 10^{-10}}$. The RJMCMC algorithm gives a similar estimate of $B_{10}=\frac{12727}{7273}\left/\left(\frac{1-\ensuremath{10^{-10}}}{\ensuremath{10^{-10}}}\right)\right.=\ensuremath{1.75\times 10^{-10}}$, meaning that model 0 is highly favoured. That the genearlised power law process hierarchical model is more appropriate for the \#gots7 data set, which consists of tweets for about 4.4 hours, is also consistent with the findings by \cite{mmnb17} for tweets collected over a similar duration of 3 hours.

\begin{knitrout}
\definecolor{shadecolor}{rgb}{0.969, 0.969, 0.969}\color{fgcolor}\begin{figure}[htbp!]

{\centering \includegraphics[width=0.4\linewidth]{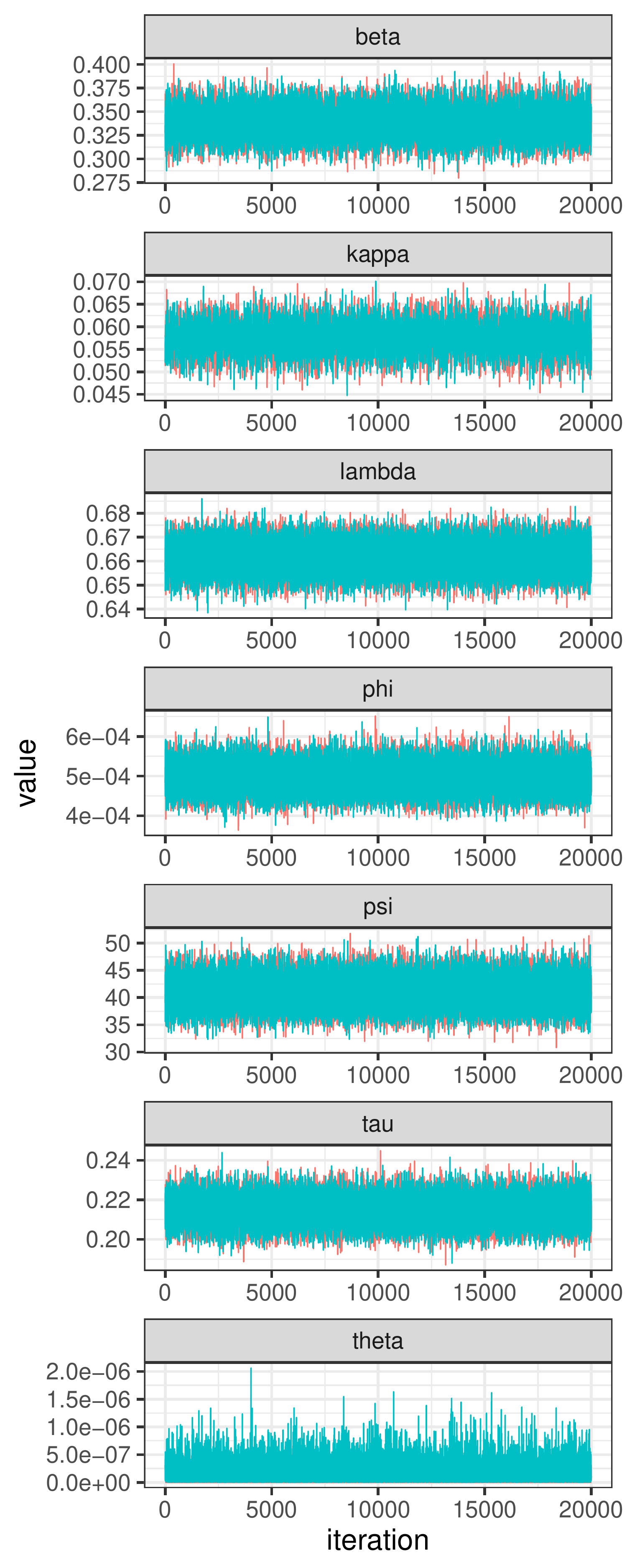} 
\includegraphics[width=0.4\linewidth]{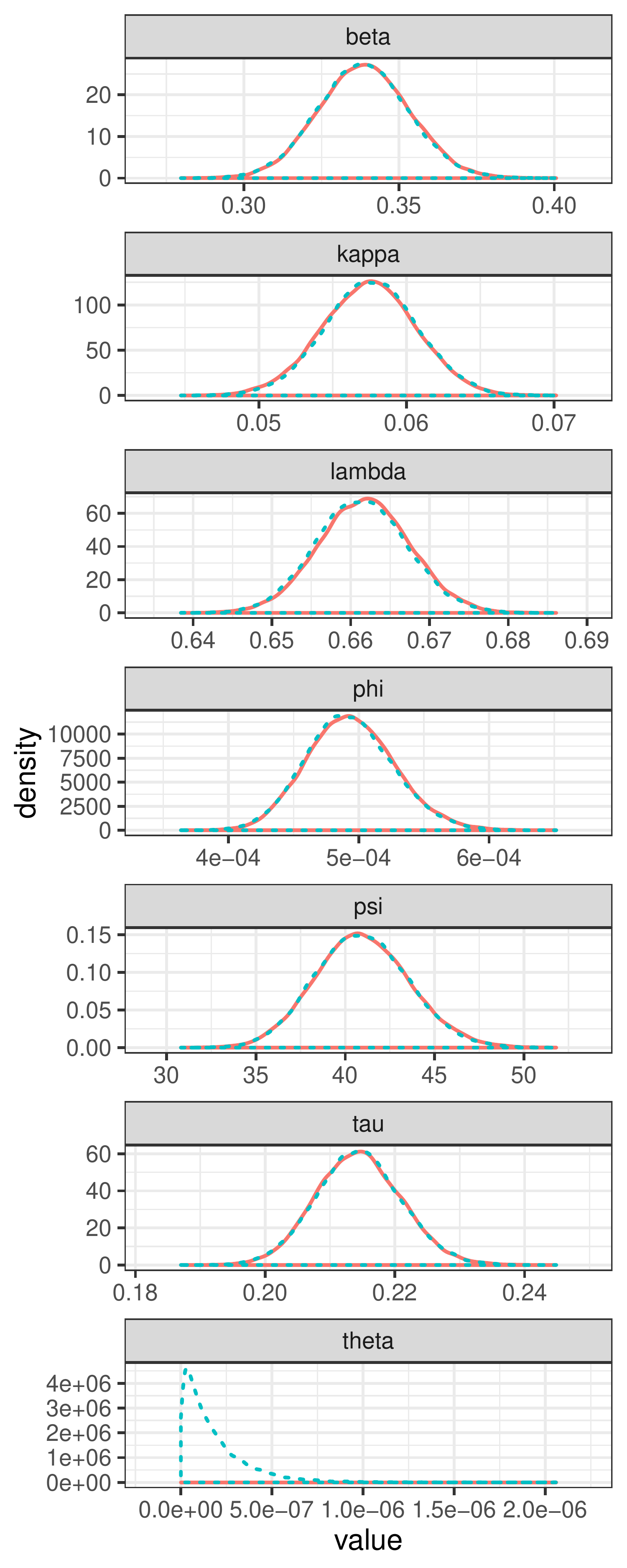} 

}

\caption[Traceplots (left) and posterior densities (right) of the parameters of models 0 (salmon, solid) and 1 (turquoise, dashed on the right) fitted to \#gots7 data]{Traceplots (left) and posterior densities (right) of the parameters of models 0 (salmon, solid) and 1 (turquoise, dashed on the right) fitted to \#gots7 data.}\label{fig.gots7_mcmc_plot}
\end{figure}

\end{knitrout}


\begin{knitrout}
\definecolor{shadecolor}{rgb}{0.969, 0.969, 0.969}\color{fgcolor}\begin{figure}[htbp!]

{\centering \includegraphics[width=0.49\linewidth]{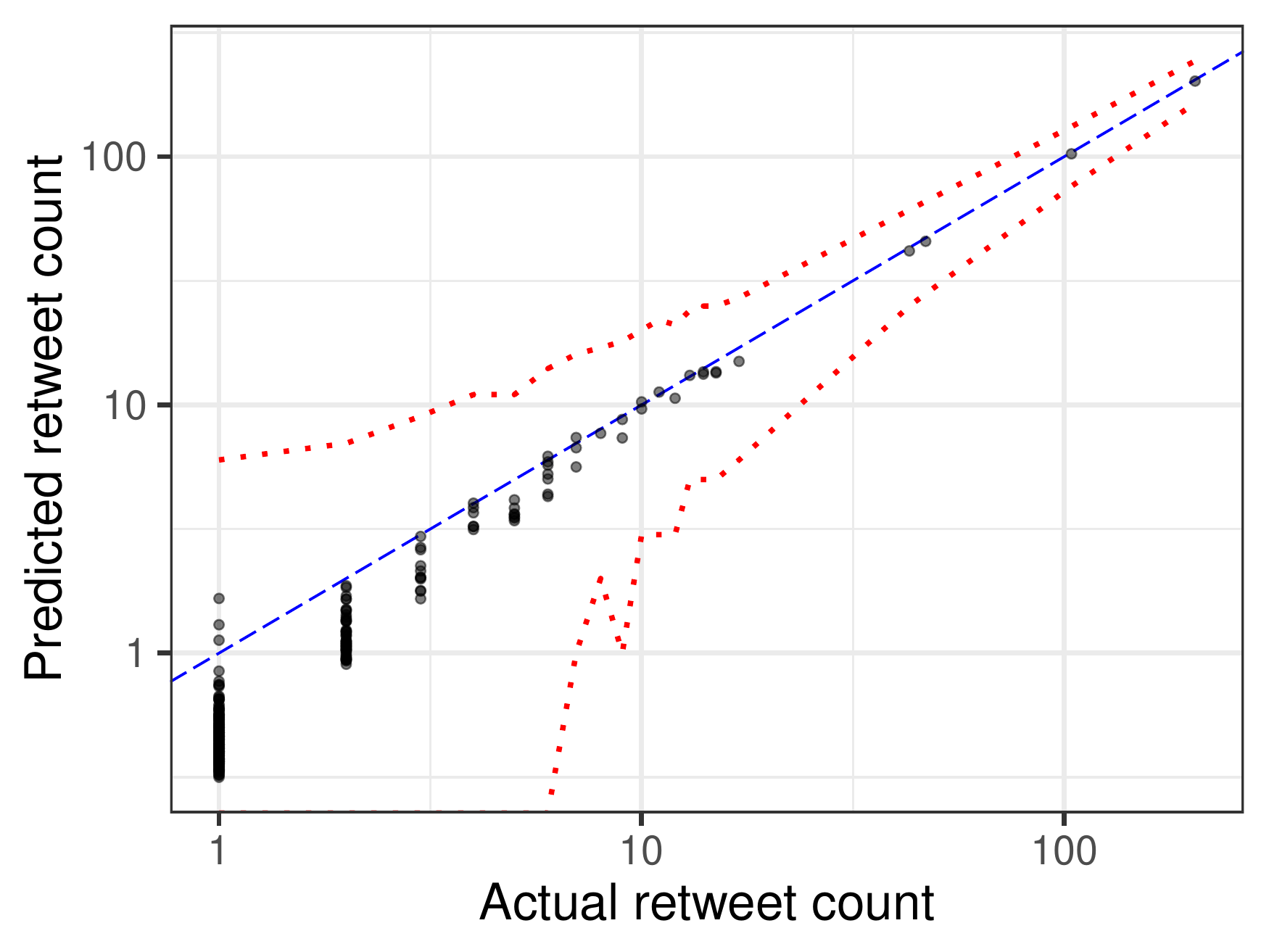} 
\includegraphics[width=0.49\linewidth]{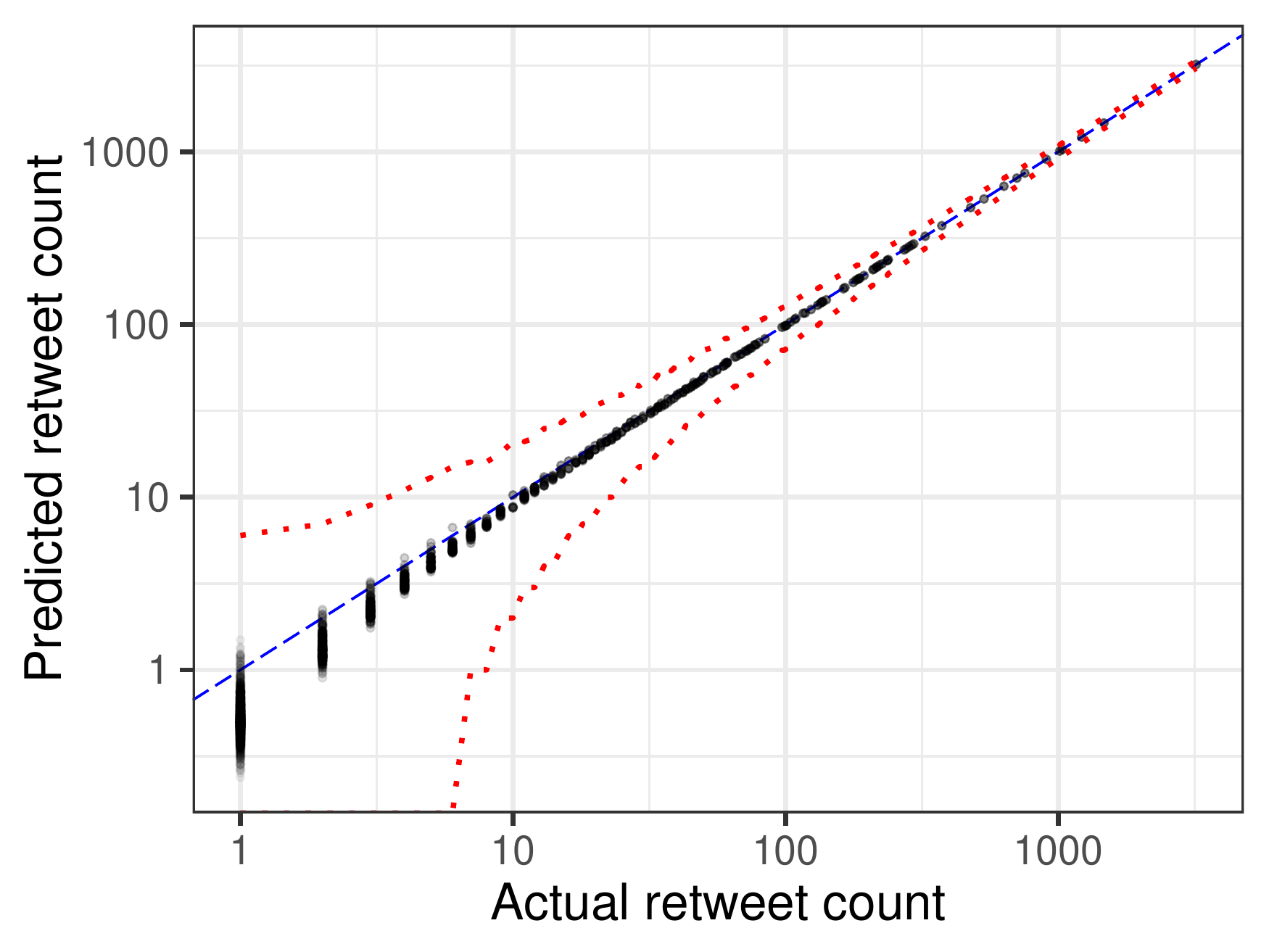} 

}

\caption[Predicted against actual retweet count for \#thehandmaidstale (left) and \#gots7 data (right) with 95\% prediction intervals (dotted, red) for the model selected by GVS]{Predicted against actual retweet count for \#thehandmaidstale (left) and \#gots7 data (right) with 95\% prediction intervals (dotted, red) for the model selected by GVS. For multiple originals with the same actual retweet count, the widest prediction interval is shown. The blue dashed line is the line $y=x$.}\label{fig.prediction}
\end{figure}

\end{knitrout}

\begin{knitrout}
\definecolor{shadecolor}{rgb}{0.969, 0.969, 0.969}\color{fgcolor}\begin{figure}[htbp!]

{\centering \includegraphics[width=0.49\linewidth]{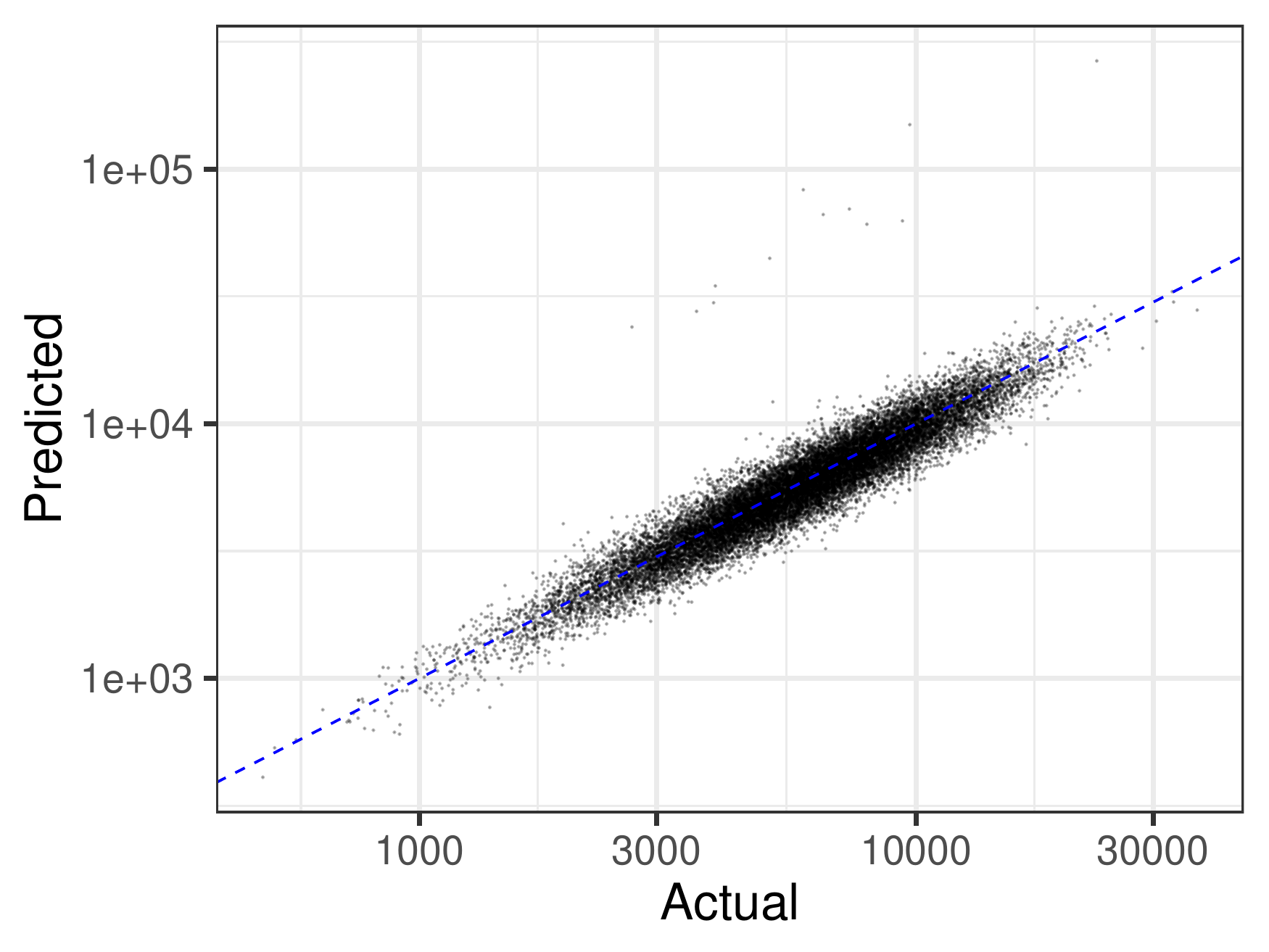} 
\includegraphics[width=0.49\linewidth]{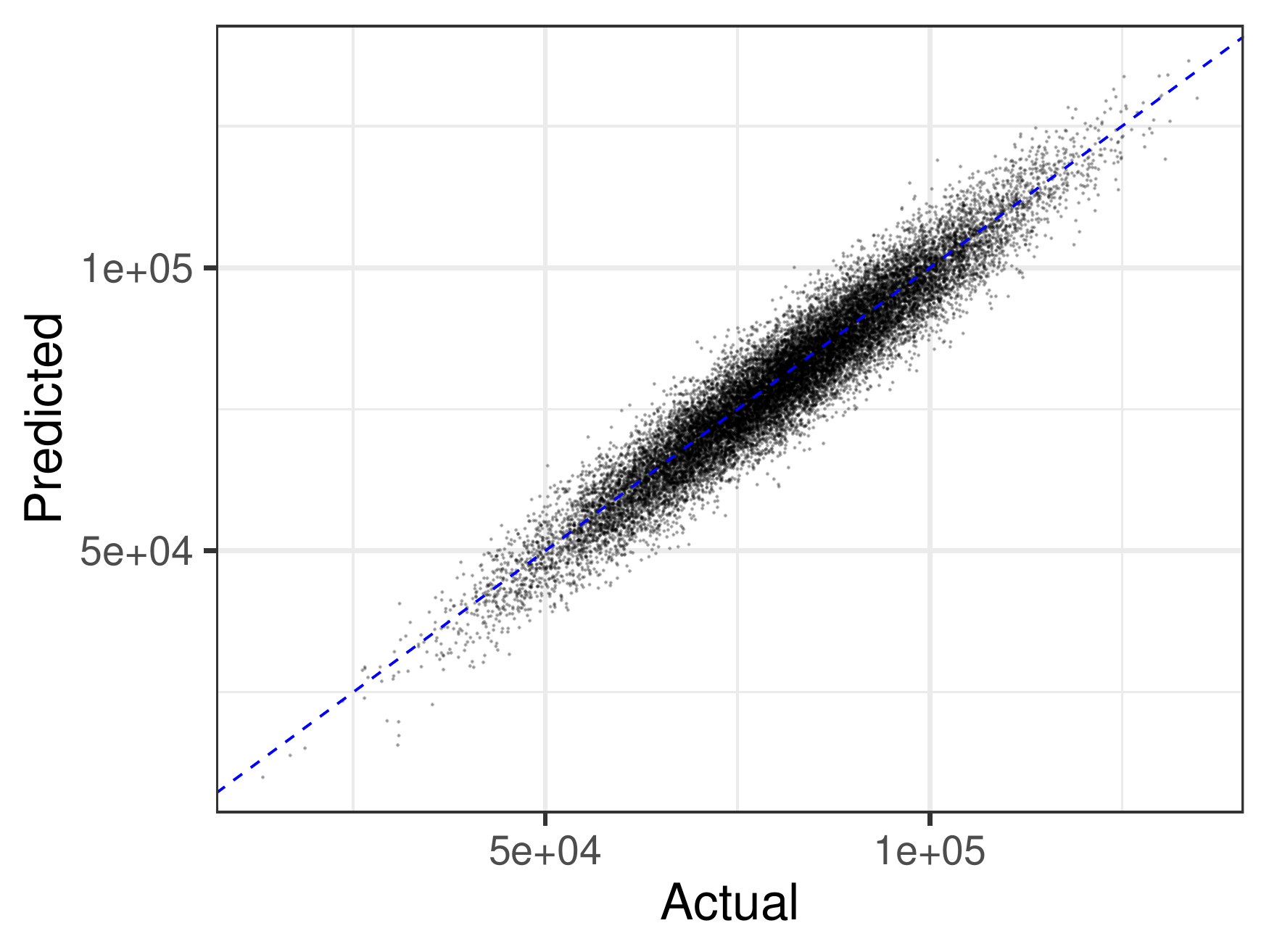} 

}

\caption[Simulated against actual $\chi^2$ discrepancies for \#thehandmaidstale (left) and \#gots7 (right) data]{Simulated against actual $\chi^2$ discrepancies for \#thehandmaidstale (left) and \#gots7 (right) data. The blue dashed line is the line $y=x$.}\label{fig.plot_chisq}
\end{figure}

\end{knitrout}

As outlined in Section \ref{sect.gof}, at each iteration of the MCMC, a value can be drawn from the Poisson distribution with mean $H_i(T;x_i,s_i,\epsilon_i,\boldsymbol{\eta})$ as the simulated (or predicted) retweet count of the $i$-th original $(i=1,2,\ldots,n)$. Across all iterations, the simulated values can be seen as a sample from the \textit{posterior predictive distribution} of the random variable of the retweet count. As a preliminary way of examining the goodness-of-fit of our model, the posterior predictive mean of the retweet count, under the selected model 0 for both data sets, is plotted against the actual retweet count in Figure \ref{fig.prediction} with 95\% predictive intervals. That some points seemingly form a vertical line is because there are multiple originals with the same actual retweet count. While small counts are slightly under-predicted, our model is doing a very good job for moderate to large counts for both data sets. For either data set, the alternative model 1 gives very close predictions of the retweet count, which are therefore not plotted, to those predicted by the selected model 0.

Utilising the simulated retweet counts, summaries of which are presented in Figure~\ref{fig.prediction}, the formal diagnostic procedures outlined in Section \ref{sect.gof} are carried out for both data sets. The simulated discrepancies $\chi^2_{\text{sim}}$ are plotted against the actual discrepancies $\chi^2_{\text{act}}$ in Figure~\ref{fig.plot_chisq}. The hovering of the points around the line $y=x$ shows no apparent lack-of-fit issues. This is supported by calculating the posterior predictive $p$-values, which are 0.424 and 0.436, for \#thehandmaidstale and \#gots7 data, respectively.

\section{Simulation and backfitting} \label{sect.sim}

For each of the two data sets, we simulate the retweet times from the NHPP with intensity \eqref{eqn.model_intensity}, based on the original times $(\boldsymbol{s})$ and the ``mean-centred'' follower counts $(\boldsymbol{x})$. The posterior means of the parameters in the respective selected models (model 0 for both data sets) are used as their true values in the simulation. The cumulative retweet counts over time for the simulated data are plotted in Figure \ref{fig.both_times_sim}. Comparing with their counterparts in Figure \ref{fig.both_times}, the simulated data show a high degree of similarity in the overall pattern of retweet growth. Furthermore, the frequencies of retweet counts are equally highly similar between the simulated data and the actual data, as shown in Figure \ref{fig.plot_counts}. Both visualisations indicate that our proposed model is very capable of generating realistic retweet times that look like the observed data.

\begin{knitrout}
\definecolor{shadecolor}{rgb}{0.969, 0.969, 0.969}\color{fgcolor}\begin{figure}[htbp!]

{\centering \includegraphics[width=0.49\linewidth]{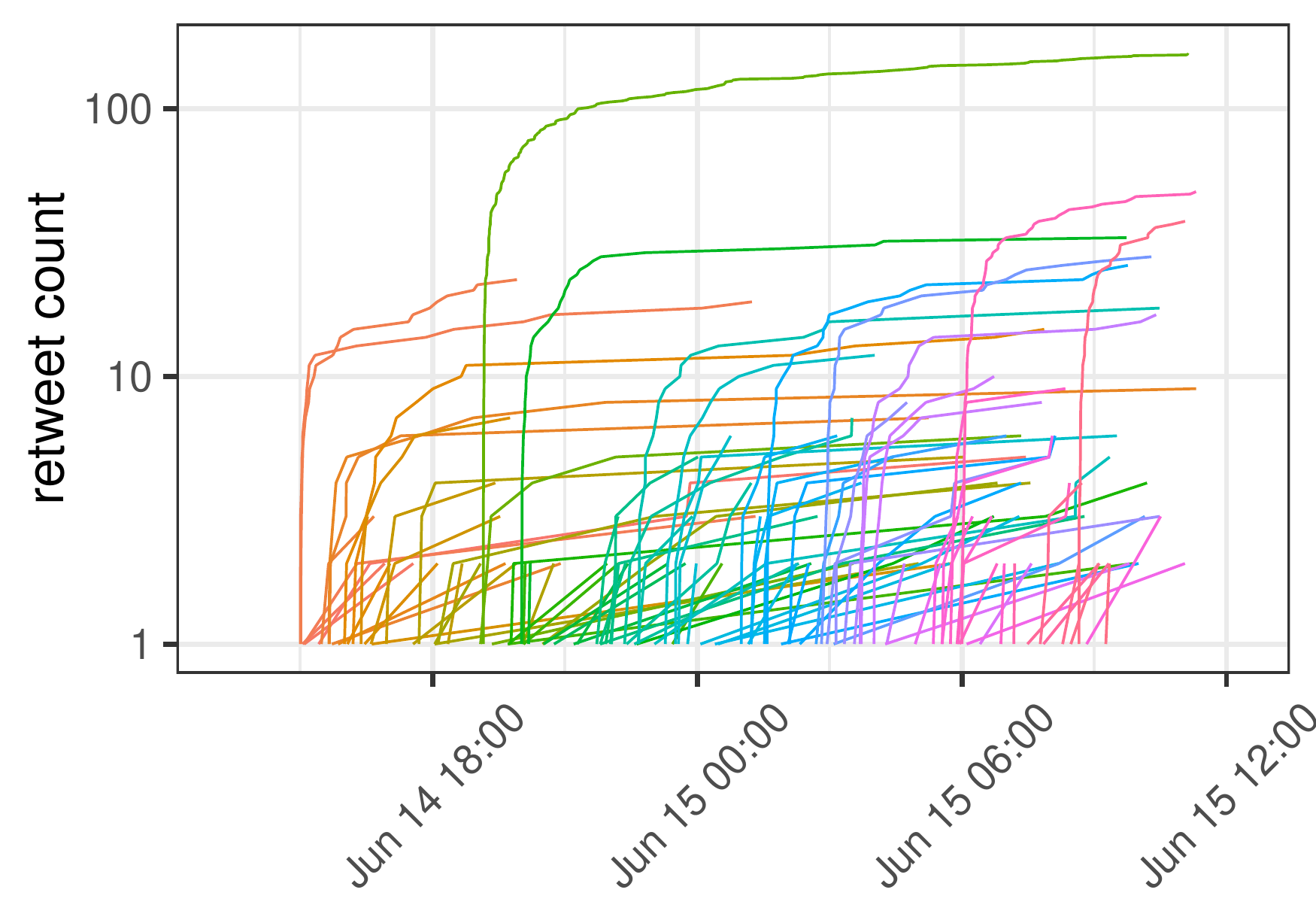} 
\includegraphics[width=0.49\linewidth]{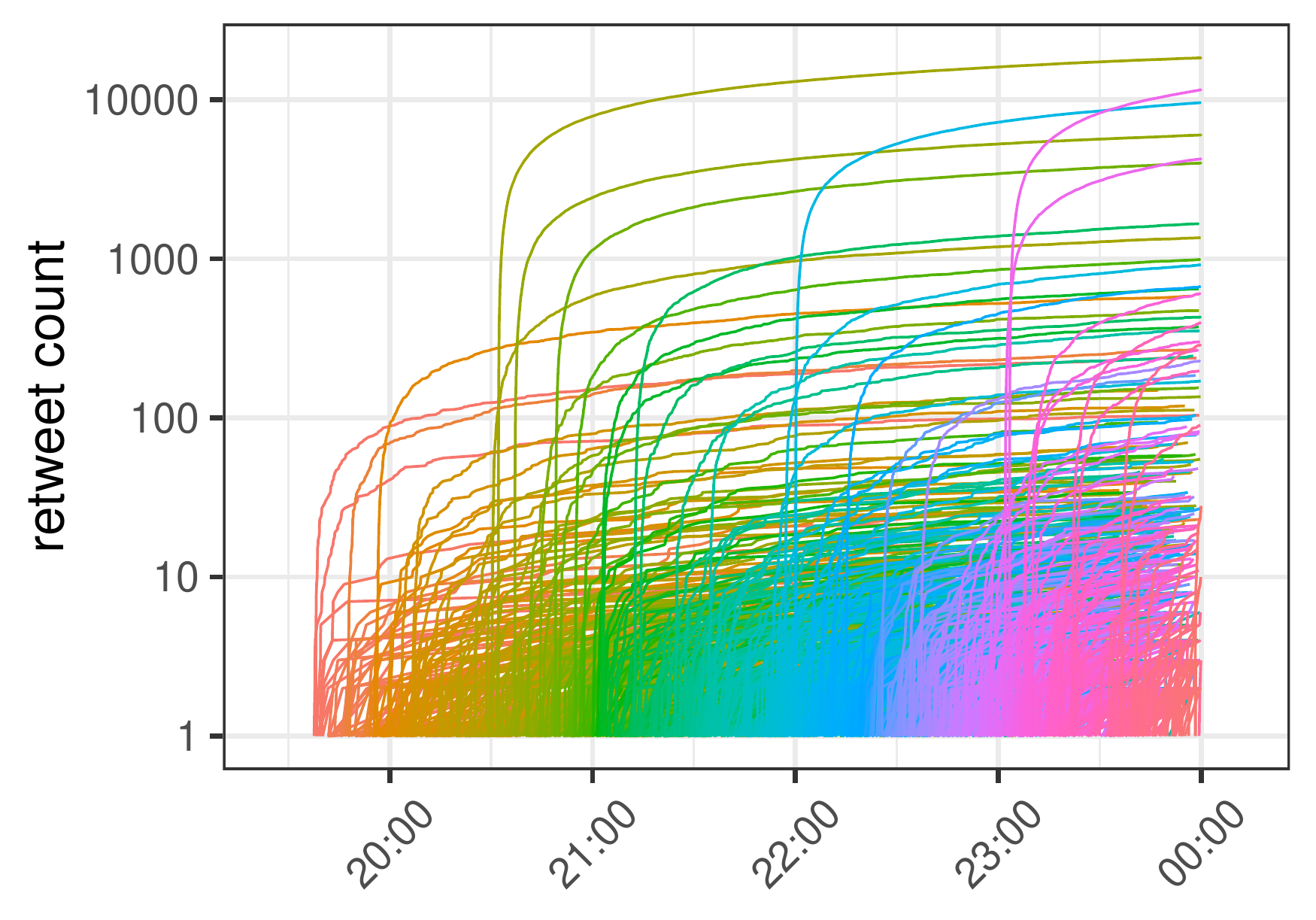} 

}

\caption[Cumulative retweet counts over time for simulated data corresponding to \#thehandmaidstale (left) and \#gots7 (right) data]{Cumulative retweet counts over time for simulated data corresponding to \#thehandmaidstale (left) and \#gots7 (right) data. The posterior means of the parameters under model 0 are used as the true values in the simulation. Each trajectory is of a different colour and represents the growth of (simulated) retweets of one individual original.}\label{fig.both_times_sim}
\end{figure}

\end{knitrout}

\begin{knitrout}
\definecolor{shadecolor}{rgb}{0.969, 0.969, 0.969}\color{fgcolor}\begin{figure}[htbp!]

{\centering \includegraphics[width=0.49\linewidth]{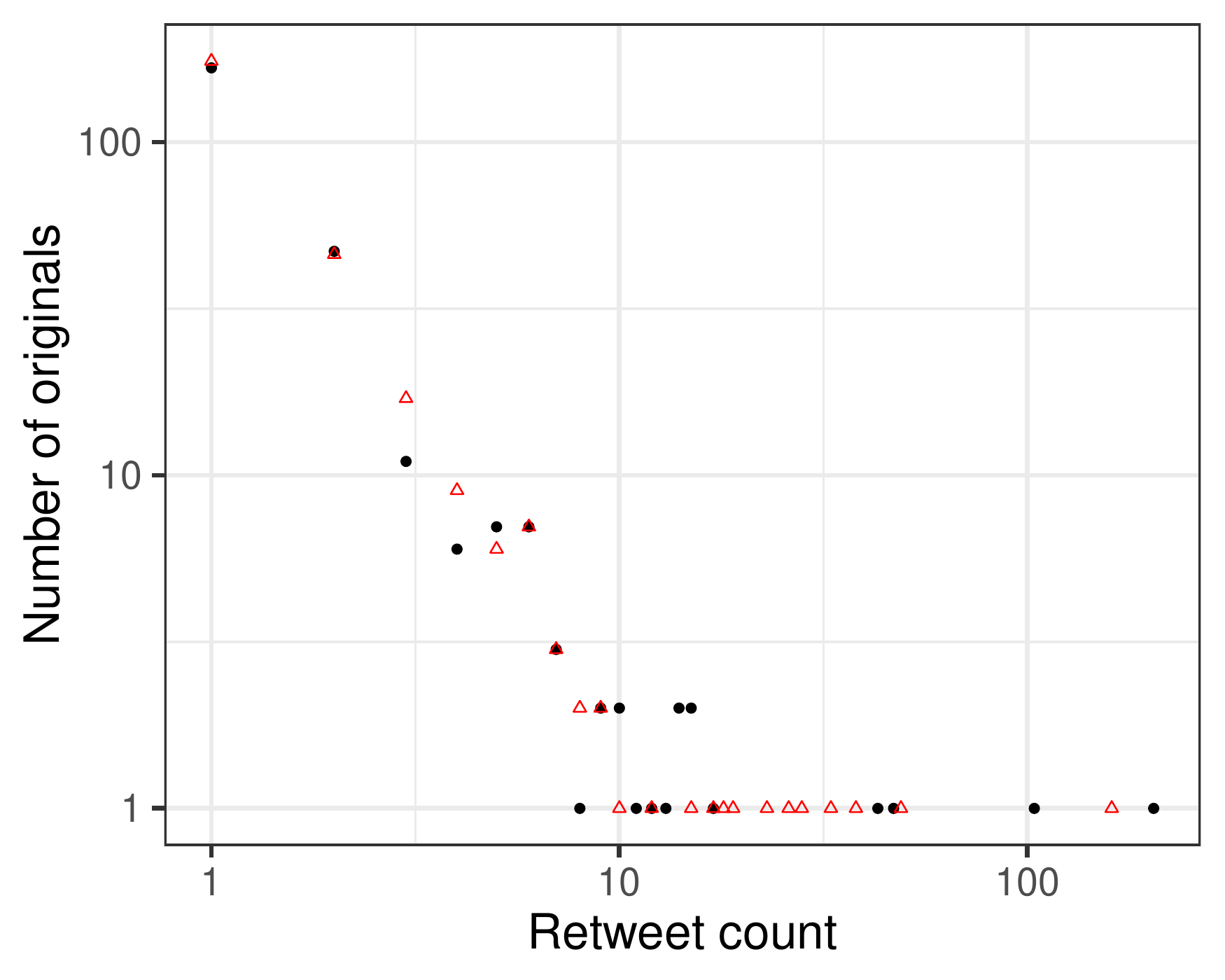} 
\includegraphics[width=0.49\linewidth]{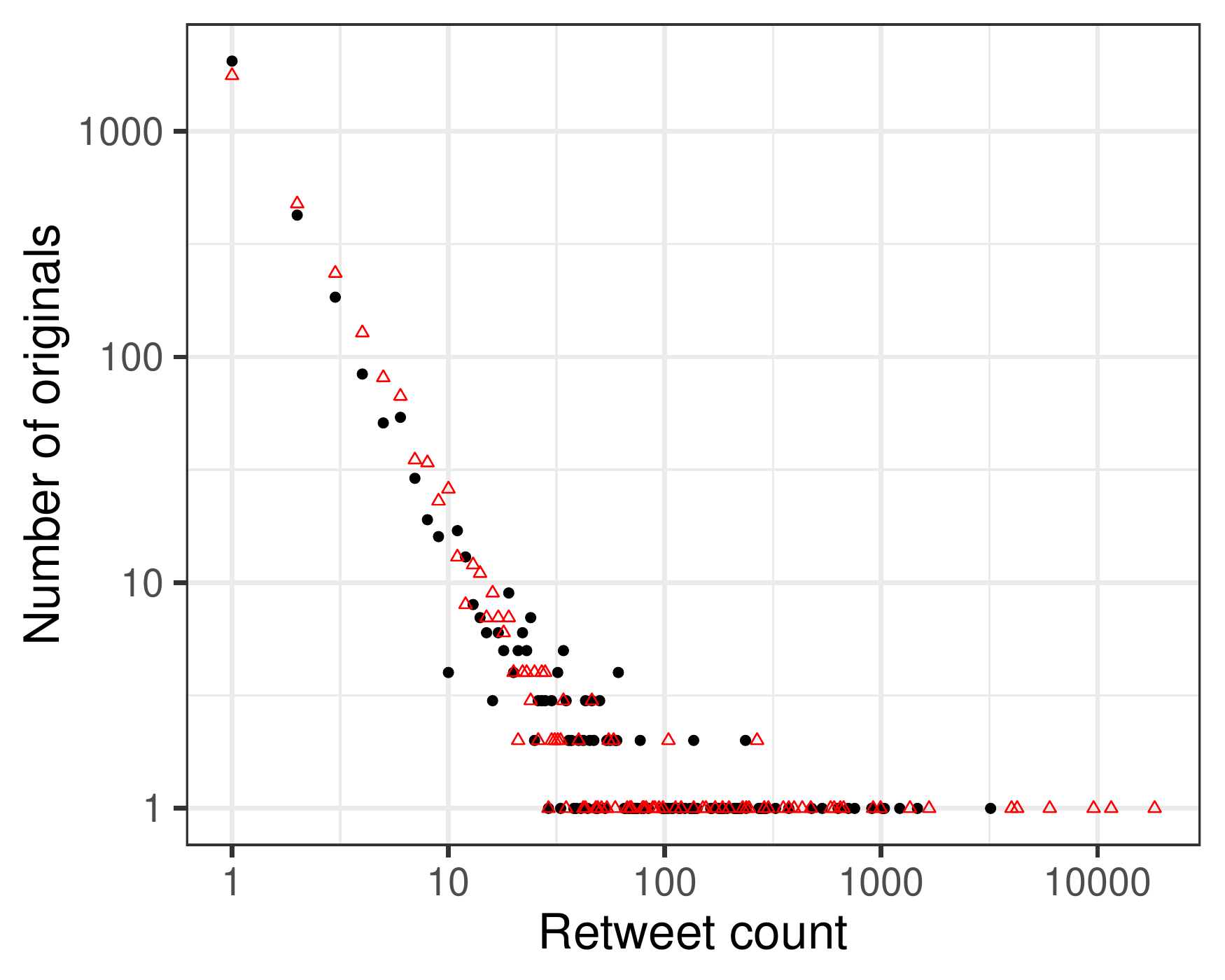} 

}

\caption[Frequencies of retweet counts for actual (black, circles) and simulated data (red, triangles) for \#thehandmaidstale (left) and \#gots7 (right) data]{Frequencies of retweet counts for actual (black, circles) and simulated data (red, triangles) for \#thehandmaidstale (left) and \#gots7 (right) data.}\label{fig.plot_counts}
\end{figure}

\end{knitrout}

\begin{knitrout}
\definecolor{shadecolor}{rgb}{0.969, 0.969, 0.969}\color{fgcolor}\begin{figure}[htbp!]

{\centering \includegraphics[width=0.96\linewidth]{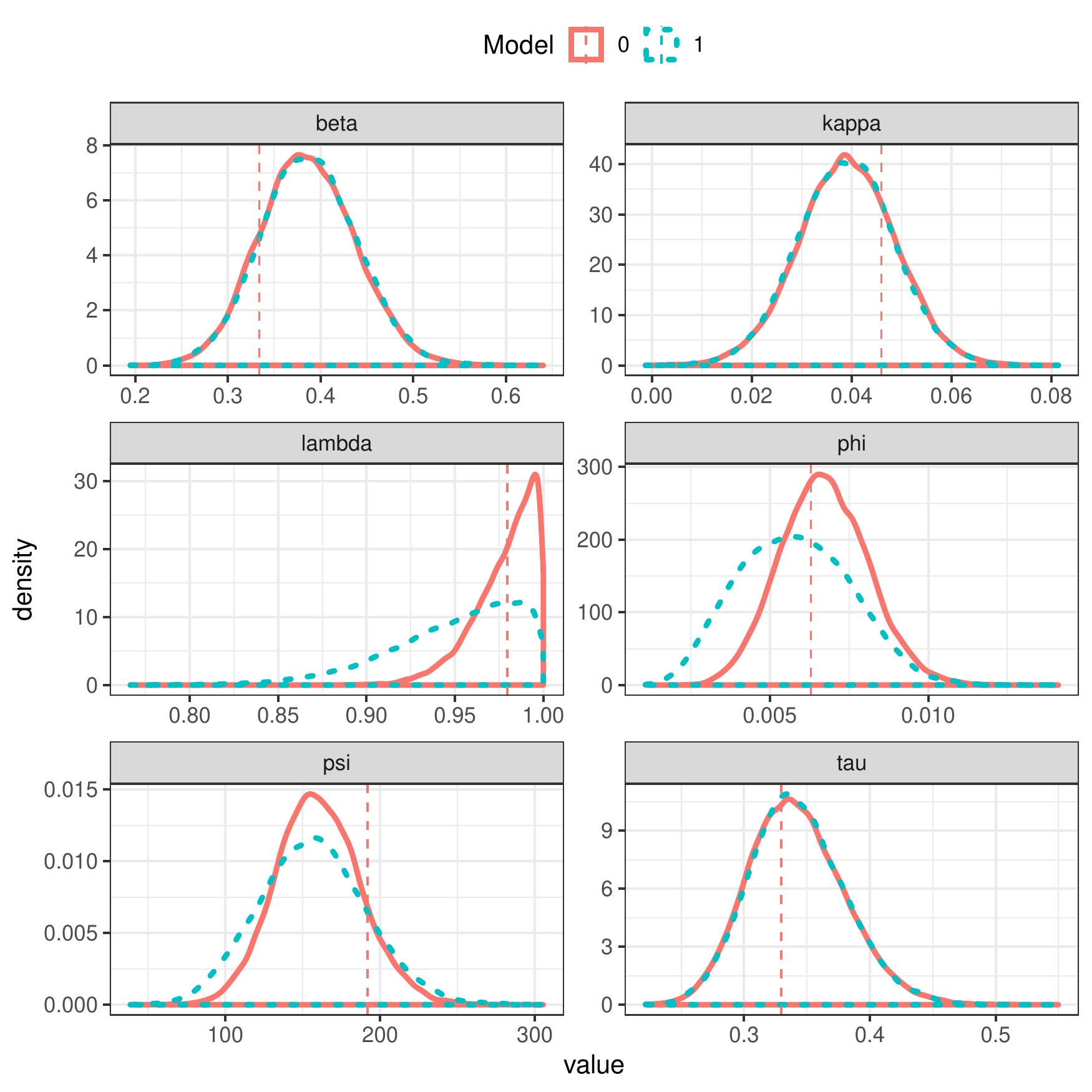} 

}

\caption{Posterior density of the parameters (except $\theta$) of models 0 (salmon, solid) and 1 (turqoise, dashed) fitted to the simulated data shown on the left of Figure \ref{fig.both_times_sim}. The vertical lines represent the respective true values of the parameters.}\label{fig.plot_sim_density_thehandmaidstale}
\end{figure}

\end{knitrout}

Not only is the model able to produce realistic realisations, but its inference procedure is also working properly and can recover the true parameter values. Specifically, we applied the MWG algorithm for the individual models (Appendix B) to the simulated data shown on the left of Figure \ref{fig.both_times_sim}. For each parameter (other than $\theta$), its true value lies within the 95\% credible interval of the corresponding marginal posterior distribution, as shown in Figure \ref{fig.plot_sim_density_thehandmaidstale}. Furthermore, the subsequent model selection overwhelming chooses model 0, which is the true model, over model 1. The Bayes factors $B_{10}$ according to GVS and RJMCMC are \ensuremath{1.829\times 10^{-8}} and \ensuremath{1.834\times 10^{-8}}, respectively.

\section{Discussion} \label{sect.con}

In this article we have proposed a Bayesian hierarchical model of generalised hybrid processes, which is shown to model well the retweets of the originals of a specific hashtag. The application to both \#thehahndmaidstale and \#gots7 data suggests that it is sufficient to fit a special case of the proposed model, which is the hierarchical model of generalised power law processes. Also, incorporating an original-specific scale $\delta_i$ allows us to explain retweet count by the follower count, which seems a natural candidate for driving retweet behaviour.

Whether the squared term $x_i^2$ should be included in $\delta_i$ can be examined by carrying out GVS or RJMCMC for $\kappa$, in the same way it is for $\theta$ in our inference and application. In fact, as GVS was originally developed for determining whether covariates in a linear regression model should be included or not, we could include terms of higher powers in the linear predictor, and perform GVS for the associated parameters (and $\kappa$ and $\theta$) simultaneously. However, as the overall fit is adequate while $0$ is excluded from the support of the marginal posterior of $\kappa$ for both data sets, and as our focus is on the overall structure of the hierarchical model instead of the particular form of the linear predictor, we confine the GVS to $\theta$ only in this article.

While there are temporal fluctuations and bursty dynamics shown by, for example, \cite{sl15}, \cite{maaj13} \cite{mm15}, for their respective Twitter data sets, our results should not been seen as contradictory, for two reasons. First, their data were collected over several weeks, during which injections of interest due to external events were possible, while the data collection period was much shorter in both of our data sets, in which the generation of tweets and retweets was predominantly due to the broadcast of a TV series episode, the time of which was pre-specified. Second, our hierarchical model concerns the behaviour of information \textit{diffusion}, in the form of retweets, which may be different from that of information \textit{generation}, in the form of originals. Had the data been collected over a much longer time span, fitting the generalised hybrid process to the originals, as we did in Section \ref{sect.data}, might not be appropriate anymore. Therefore, even though the generalised hybrid process may not be applicable to all kinds of online social media data of all timescales, it should still be useful to data regarding information diffusion within a time span of a single event with no apparent changepoints.

The parameters $\beta,\kappa,\lambda,\phi,\psi,\tau$ and $\theta$ are assumed to be common to all generalised hybrid processes of the originals. We argue that, for $\beta,\kappa,\phi$ and $\tau$, the between-original variation is already captured by the covariate $x_i$ together with $\epsilon_i$, while for $\psi$ there is simply no extra information to assume otherwise. It might however be useful to incorporate a hierarchical structure to $\lambda$ and $\theta$ simultaneously, to allow variation in the rate of decay of information diffusion between originals of the same hashtag. The parsimony of the proposed model also makes it easy to simulate retweets given originals and follower counts. Given the parameters, we can directly simulate $\epsilon_i$, $\delta_i$ and subsequently the retweet times using \eqref{eqn.model_e}, \eqref{eqn.model_delta} and \eqref{eqn.model_intensity}, respectively. If appropriate, we can go one step further by first simulating the originals from a more general point process, followed by the simulation of the retweets.

Another direction for further study is the generation of follower count. While both the follower count and the retweet count in our data sets empirically follow the power law, it is not used to characterise either of them. Instead of treating the former as a given covariate, it is possible to first draw the follower count from the power law, then draw retweet count given the follower count, via the implicit Poisson regression model.

\if1\blind {
  \section*{Acknowledgements}
  Data supporting this publication is openly available under an `Open Data Commons Open Database License'. Additional metadata are available at: http://dx.doi.org/10.17634/154300-57. Please contact Newcastle Research Data Service at rdm@ncl.ac.uk for access instructions. This research was funded by the Engineering and Physical Sciences Research Council (EPSRC) grant DERC: Digital Economy Research Centre (EP/M023001/1). 
} \fi

\bibliographystyle{agsm}
\bibliography{ref_ca}
\appendix
\section{Likelihood derivation} \label{sect.appendix_lik}
This appendix derives the likelihood of the hierarchical model for retweets in Section 3. We first consider the contribution of the $i$-th original regardless of the model choice, or the value of $M$ equivalently. The contribution to the likelihood is
\begin{align}
  f(m_i,\boldsymbol{t}_i|x_i,s_i,\epsilon_i,\boldsymbol{\eta}_{M},M)=\exp\left(-H_i(T)\right)\times\left\{\begin{array}{ll}
  1, & m_i=0,\\
  \displaystyle\prod_{j=1}^{m_i}h_i\left(t_{ij}\right), & m_i>0.
  \end{array}\right.\nonumber
\end{align}
The general expression above is essentially the same as (2). The complete likelihood is
\begin{align}
  &f(\boldsymbol{m},\boldsymbol{t}|\boldsymbol{x},\boldsymbol{s},\boldsymbol{\epsilon},\boldsymbol{\eta}_{M},M)
  =\mathlarger\prod_{i=1}^{n}f(m_i,\boldsymbol{t}_i|x_i,s_i,\epsilon_i,\boldsymbol{\eta}_{M},M)\nonumber\\
  &\qquad=\mathlarger\prod_{i:m_i=0}^{}\exp\left(-H_i(T)\right)\times\mathlarger\prod_{i:m_i>0}^{}\left[\exp\left(-H_i(T)\right)\prod_{j=1}^{m_i}h_i(t_{ij})\right]\nonumber\\
  &\qquad=\mathlarger\prod_{i=1}^{n}\exp\left(-H_i(T)\right)\times\mathlarger\prod_{i:m_i>0}^{}\mathlarger\prod_{j=1}^{m_{i}}h_{i}(t_{ij}).\tag{*}\label{eqn.appendix_lik}
\end{align}
When $M=0$, that is, $\theta=0$ and is removed from the parameter vector, substituting (5) and (6) into \eqref{eqn.appendix_lik} yields
\begin{align}
  &f(\boldsymbol{m},\boldsymbol{t}|\boldsymbol{x},\boldsymbol{s},\boldsymbol{\epsilon},\boldsymbol{\eta}_0,M=0)\nonumber\\*
  &\qquad=\mathlarger\prod_{i=1}^{n}\exp\left(-\phi~e^{\DELTA}\left[\left(T-s_i+\psi\right)^{1-\lambda}-\psi^{1-\lambda}\right]\left(1-\lambda\right)^{-1}\right)\nonumber\\
  &\qquad\quad\times\mathlarger\prod_{i:m_i>0}^{}\mathlarger\prod_{j=1}^{m_i}\left[\phi~e^{\DELTA}\left(t_{ij}-s_i+\psi\right)^{-\lambda}\right]\nonumber\\
  &\qquad=\exp\left(-\sum_{i=1}^n\phi~e^{\DELTA}\left[\left(T-s_i+\psi\right)^{1-\lambda}-\psi^{1-\lambda}\right]\left(1-\lambda\right)^{-1}\right)\nonumber\\
  &\qquad\quad\times\mathlarger\prod_{i:m_i>0}^{}\left[\phi^{m_i}\left(e^{\DELTA}\right)^{m_i}\mathlarger\prod_{j=1}^{m_{i}}\left(t_{ij}-s_i+\psi\right)^{-\lambda}\right]\nonumber\\
  &\qquad=\exp\left(-\frac{\phi}{1-\lambda}\sum_{i=1}^ne^{\DELTA}\left[\left(T-s_i+\psi\right)^{1-\lambda}-\psi^{1-\lambda}\right]\right)\nonumber\\
  &\qquad\quad\times\phi^{\sum_{i:m_i>0}m_i}\exp\left(\sum_{i:m_i>0}^{}m_i\left[\DELTA\right]\right)\times\mathlarger\prod_{i:m_i>0}\mathlarger\prod_{j=1}^{m_{i}}\left(t_{ij}-s_i+\psi\right)^{-\lambda},\nonumber\\
  &\qquad=\exp\left(-\frac{\phi}{1-\lambda}\sum_{i=1}^ne^{\DELTA}\left[\left(T-s_i+\psi\right)^{1-\lambda}-\psi^{1-\lambda}\right]\right)\nonumber\\
  &\qquad\quad\times\phi^{\sum_{i=1}^{n}m_i}\exp\left(\sum_{i=1}^{n}m_i\left[\DELTA\right]\right)\times\mathlarger\prod_{i:m_i>0}\mathlarger\prod_{j=1}^{m_{i}}\left(t_{ij}-s_i+\psi\right)^{-\lambda},\nonumber
\end{align}
which is identical to (11) as $\displaystyle\sum_{i=1}^nm_i=m$. When $M=1$, that is, $\theta>0$, \eqref{eqn.appendix_lik} becomes
\begin{align}
  &f(\boldsymbol{m},\boldsymbol{t}|\boldsymbol{x},\boldsymbol{s},\boldsymbol{\epsilon},\boldsymbol{\eta}_1,M=1)\nonumber\\*
  &\qquad=\mathlarger\prod_{i=1}^{n}\exp\left(-\phi~e^{\DELTA}\left[\Gamma(1-\lambda,\theta\left(T-s_i+\psi\right))-\Gamma(1-\lambda,\theta\psi)\right]\theta^{\lambda-1}e^{\theta\psi}\right),\nonumber\\
  &\qquad\quad\times\mathlarger\prod_{i:m_i>0}\mathlarger\prod_{j=1}^{m_i}\left[\phi~e^{\DELTA}\left(t_{ij}-s_i+\psi\right)^{-\lambda}e^{-\theta(t_{ij}-s_i)}\right]\nonumber\\
  &\qquad=\exp\left(-\sum_{i=1}^{n}\phi~e^{\DELTA}\left[\Gamma(1-\lambda,\theta\left(T-s_i+\psi\right))-\Gamma(1-\lambda,\theta\psi)\right]\theta^{\lambda-1}e^{\theta\psi}\right)\nonumber\\
  &\qquad\quad\times\mathlarger\prod_{i:m_i>0}\left[\phi^{m_i}\left(e^{\DELTA}\right)^{m_i}\mathlarger\prod_{j=1}^{m_{i}}\left(t_{ij}-s_i+\psi\right)^{-\lambda}\mathlarger\prod_{j=1}^{m_{i}}e^{-\theta(t_{ij}-s_i)}\right]\nonumber\\
  &\qquad=\exp\left(-\phi~\theta^{\lambda-1}e^{\theta\psi}\sum_{i=1}^{n}e^{\DELTA}\left[\Gamma(1-\lambda,\theta\left(T-s_i+\psi\right))-\Gamma(1-\lambda,\theta\psi)\right]\right)\nonumber\\
  &\qquad\quad\times\phi^{\sum_{i:m_i>0}m_i}\exp\left(\sum_{i:m_i>0}m_i\left[\DELTA\right]\right)\nonumber\\
  &\qquad\quad\times\mathlarger\prod_{i:m_i>0}\mathlarger\prod_{j=1}^{m_i}\left(t_{ij}-s_i+\psi\right)^{-\lambda}\times\mathlarger\prod_{i:m_i>0}\mathlarger\prod_{j=1}^{m_i}e^{-\theta(t_{ij}-s_i)}\nonumber\\
  &\qquad=\exp\left(-\phi~\theta^{\lambda-1}e^{\theta\psi}\sum_{i=1}^{n}e^{\DELTA}\left[\Gamma(1-\lambda,\theta\left(T-s_i+\psi\right))-\Gamma(1-\lambda,\theta\psi)\right]\right)\nonumber\\
  &\qquad\quad\times\phi^{\sum_{i=1}^nm_i}\exp\left(\sum_{i=1}^nm_i\left[\DELTA\right]\right)\nonumber\\
  &\qquad\quad\times\mathlarger\prod_{i:m_i>0}\mathlarger\prod_{j=1}^{m_i}\left(t_{ij}-s_i+\psi\right)^{-\lambda}\times\exp\left(-\theta\sum_{i:m_i>0}\sum_{j=1}^{m_i}\left(t_{ij}-s_i\right)\right),\nonumber
\end{align}
which is identical to (12).

\section{MWG algorithm} \label{sect.appendix_mcmc}
This appendix describes the MWG algorithm for $\boldsymbol{\eta}_{M}$ and $\boldsymbol{\epsilon}$ in steps 2 and 3 of the algorithm outlined in Section 4. For any scalar parameter $\sigma$, we define $\boldsymbol{\eta}_{M,-\sigma}$ to be the vector $\boldsymbol{\eta}_{M}$ without $\sigma$, and for notational convenience, we denote $\boldsymbol{\eta}_{M}$ with $\sigma$ set to a value $\sigma_0$ by $(\sigma_0,\boldsymbol{\eta}_{M,-\sigma})$, regardless of the position of $\sigma$ in $\boldsymbol{\eta}_{M}$. 

\textbf{Sampling $\beta$:} Assume the current value is $\beta$. We propose $\beta^{*}$ from a symmetrical proposal $q(\cdot|\beta)$, and accept $\beta^{*}$ with probability 
\begin{align}
  \min\left(1,\frac
  {f\left(\boldsymbol{m},\boldsymbol{t}|\boldsymbol{x},\boldsymbol{s},\boldsymbol{\epsilon},(\beta^{*},\boldsymbol{\eta}_{M,-\beta}),M\right)\pi_{\beta}(\beta^{*})}
  {f\left(\boldsymbol{m},\boldsymbol{t}|\boldsymbol{x},\boldsymbol{s},\boldsymbol{\epsilon},(\beta^{}, \boldsymbol{\eta}_{M,-\beta}),M\right)\pi_{\beta}(\beta^{})}
  \right),\nonumber
\end{align}
where $\pi_{\beta}(\cdot)$, along with the priors for other scalars, is given by (13).

\textbf{Sampling $\kappa$:} Assume the current value is $\kappa$. We propose $\kappa^{*}$ from a symmetrical proposal $q(\cdot|\kappa)$, and accept $\kappa^{*}$ with probability
\begin{align}
  \min\left(1,\frac
  {f\left(\boldsymbol{m},\boldsymbol{t}|\boldsymbol{x},\boldsymbol{s},\boldsymbol{\epsilon},(\kappa^{*},\boldsymbol{\eta}_{M,-\kappa}),M\right)\pi_{\kappa}(\kappa^{*})}
  {f\left(\boldsymbol{m},\boldsymbol{t}|\boldsymbol{x},\boldsymbol{s},\boldsymbol{\epsilon},(\kappa^{}, \boldsymbol{\eta}_{M,-\kappa}),M\right)\pi_{\kappa}(\kappa^{})}
  \right).\nonumber
\end{align}

\textbf{Sampling $\lambda$:} Assume the current value is $\lambda^{}$. We propose $\lambda^{*}$ from a symmetrical proposal $q(\cdot|\lambda)$, and accept $\lambda^{*}$ with probability
\begin{align}
  \min\left(1,\frac
  {f\left(\boldsymbol{m},\boldsymbol{t}|\boldsymbol{x},\boldsymbol{s},\boldsymbol{\epsilon},(\lambda^{*},\boldsymbol{\eta}_{M,-\lambda}),M\right)\pi_{\lambda}(\lambda^{*})\boldsymbol{1}_{\{\lambda^{*}<1\}}}
  {f\left(\boldsymbol{m},\boldsymbol{t}|\boldsymbol{x},\boldsymbol{s},\boldsymbol{\epsilon},(\lambda^{}, \boldsymbol{\eta}_{M,-\lambda}),M\right)\pi_{\lambda}(\lambda^{}) \boldsymbol{1}_{\{\lambda^{}<1\}}}
  \right).\nonumber
\end{align}

\textbf{Sampling $\phi$:} As we have assigned a conditional conjugate prior to $\phi$, its full conditional posterior is given by $\phi|\boldsymbol{x},\boldsymbol{s},\boldsymbol{\epsilon},\boldsymbol{m},\boldsymbol{t},\boldsymbol{\eta}_{M,-\phi}\sim\text{Gamma}\left(a_{\phi}^{*},b_{\phi}^{*}\right)$, where $a_{\phi}^{*}=a_{\phi}+m$ and
\begin{align}
  b_{\phi}^{*}=b_{\phi}+\left\{\begin{array}{ll}
  \displaystyle\frac{1}{1-\lambda}\sum_{i=1}^{n}e^{\delta_i}\left[\left(T-s_i+\psi\right)^{1-\lambda}-\psi^{1-\lambda}\right], & \theta = 0,\\
  \displaystyle\theta^{\lambda-1}e^{\theta\psi}\sum_{i=1}^{n}e^{\delta_i}\left[\Gamma\left(1-\lambda,\theta\left(T-s_i+\psi\right)\right)-\Gamma\left(1-\lambda,\theta\psi\right)\right], & \theta > 0,
  \end{array}\right.\nonumber
\end{align}
and $\delta_i=\DELTA$ as in (7).

\textbf{Sampling $\psi$:} Assume the current value is $\psi$. We propose $\psi^{*}$ from a symmetrical proposal $q(\cdot|\psi)$, and accept $\psi^{*}$ with probability
\begin{align}
  \min\left(1,\frac
  {f\left(\boldsymbol{m},\boldsymbol{t}|\boldsymbol{x},\boldsymbol{s},\boldsymbol{\epsilon},(\psi^{*},\boldsymbol{\eta}_{M,-\psi}),M\right)\pi_{\psi}(\psi^{*})\boldsymbol{1}_{\{\psi^{*}\geq0\}}}
  {f\left(\boldsymbol{m},\boldsymbol{t}|\boldsymbol{x},\boldsymbol{s},\boldsymbol{\epsilon},(\psi^{},\boldsymbol{\eta}_{M,-\psi}),M\right)\pi_{\psi}(\psi^{})\boldsymbol{1}_{\{\psi^{*}\geq0\}}}
  \right).\nonumber
\end{align}

\textbf{Sampling $\tau$:} As we have assigned a conditional conjugate prior to $\tau$, its full conditional posterior is given by
\begin{align}
  \tau|\boldsymbol{x},\boldsymbol{s},\boldsymbol{\epsilon},\boldsymbol{m},\boldsymbol{t},\boldsymbol{\eta}_{M,-\tau}\sim\text{Gamma}\left(a_{\tau}+\frac{n}{2},b_{\tau}+\frac{1}{2}\sum_{i=1}^{n}\epsilon_i^2\right)
  .\nonumber
\end{align}
Effectively, we can sample for $\tau$ via a Gibbs step conditional on just the latent variables $\boldsymbol{\epsilon}$.

\textbf{Sampling $\epsilon_i~(i=1,2,\ldots,n)$:} Assume the current value is $\epsilon_i^{}$. We propose $\epsilon^{*}_i$ from a symmetrical proposal $q(.|\epsilon_i)$, and accept $\epsilon^{*}_i$ with probability
\begin{align}
  \min\left(1,\frac
  {\exp\left[-H_i\left(T;x_i,s_i,\epsilon_i^{*},\boldsymbol{\eta}\right)+m_i\epsilon_i^{*}\right]\pi_{\epsilon}(\epsilon^{*}_i|\tau)}
  {\exp\left[-H_i\left(T;x_i,s_i,\epsilon_i,\boldsymbol{\eta}\right)+m_i\epsilon_i\right]\pi_{\epsilon}(\epsilon^{}_i|\tau)}
  \right),\nonumber
\end{align}
where the notation and expression of $H_i(T;x_i,s_i,\epsilon_i,\boldsymbol{\eta})$ are given by (16) and (6), respectively.

\textbf{Sampling $\theta$ (for $M=1$ only):} Assume the current value is $\theta^{}$. We propose $\theta^{*}$ from a symmetrical proposal $q(\cdot|\theta)$, and accept $\theta^{*}$ with probability
\begin{align}
  \min\left(1,\frac
  {f\left(\boldsymbol{m},\boldsymbol{t}|\boldsymbol{x},\boldsymbol{s},\boldsymbol{\epsilon},(\theta^{*},\boldsymbol{\eta}_{1,-\theta}),M=1\right)\pi_{\theta}(\theta^{*})\boldsymbol{1}_{\{\theta^{*}>0\}}}
  {f\left(\boldsymbol{m},\boldsymbol{t}|\boldsymbol{x},\boldsymbol{s},\boldsymbol{\epsilon},(\theta^{}, \boldsymbol{\eta}_{1,-\theta}),M=1\right)\pi_{\theta}(\theta^{})\boldsymbol{1}_{\{\theta^{}>0\}}}
  \right).\nonumber
\end{align}

\section{RJMCMC algorithm} \label{sect.appendix_rjmcmc}

This appendix describes the algorithm of RJMCMC, which is an alternative to GVS for model selection outlined in Section 4. In additional to the notation defined in previous sections, we denote $p(m,m^{'})$ as the jump probability from model $m$ to model $m^{'}$. This probability can be chosen to optimise the mixing of the algorithm, and has to be pre-specified for all pairs of $m$ and $m^{'}$ (including $m=m^{'}$). As the posterior model probabilities are theoretically not affected by the jump probabilities, in our application $p(0,1)$ and $p(1,0)$ are chosen to both be 0.5 for simplicity. Both $p(0,0)$ and $p(1,1)$ are subsequently 0.5 too, as $p(0,0)+p(0,1)=p(1,0)+p(1,1)=1$. The algorithm is as follows:

\begin{enumerate}
  \item The current values in the chain are $\boldsymbol{\epsilon}$, $\boldsymbol{\eta}_{M}, \boldsymbol{\eta}_{\backslash M}$ and $M$.
  \item Propose a jump to models $M$ and $1-M$ with probabilities $p(M,M)$ and $p(M,1-M)$, respectively.
  \item If it is model $M$ the jump is proposed to, update $\boldsymbol{\epsilon}$ and $\boldsymbol{\eta}_M$ using the MWG algorithm described in Appendix \ref{sect.appendix_mcmc}, and the current value of $M$ stays unchanged. If it is model $1-M$ the jump is proposed to, go to the next two steps.
  \item If $M=0$ (and it is $M=1$ the jump is proposed to), draw $\theta$ from its pseudoprior $\pi_{\theta}(\theta|M=0)$, and write $\boldsymbol{\eta}_1^{'}=(\boldsymbol{\eta}_0,\theta)$. If $M=1$ (and it is $M=0$ the jump is proposed to), write $\boldsymbol{\eta}_0^{'}=\boldsymbol{\eta}_{1,-\theta}$, that is, $\boldsymbol{\eta}_1$ with the value of $\theta$ dropped, so that $\boldsymbol{\eta}_1=(\boldsymbol{\eta}_0^{'},\theta)$.
  \item Accept the proposed move to $\boldsymbol{\eta}_{1-M}^{'}$ and model $1-M$ with probability
\end{enumerate}
\begin{align}
  \min\left(1,\frac
           {f(\boldsymbol{m},\boldsymbol{t}|\boldsymbol{x},\boldsymbol{s},\boldsymbol{\epsilon},\boldsymbol{\eta}_{1-M}^{'},1-M)~\pi_{\theta}(\theta|1-M)~\pi(1-M)~p(1-M,M)}
           {f(\boldsymbol{m},\boldsymbol{t}|\boldsymbol{x},\boldsymbol{s},\boldsymbol{\epsilon},\boldsymbol{\eta}_M,M)~\pi_{\theta}(\theta|M)~\pi(M)~p(M,1-M)}
           \right).\nonumber
\end{align}

As in the GVS algorithm, both the pseudoprior $\pi_{\theta}(\theta|M=0)$ and the prior $\pi_{\theta}(\theta|M=1)$ of $\theta$ are involved in calculating the above probability, while the priors of the overlapping parameters are not required as they are the same under both models.

\end{document}